\documentclass[]{elsarticle}

\usepackage[titletoc]{appendix}
\usepackage{supertabular}
\usepackage{lineno,hyperref}
\usepackage{siunitx}
\usepackage{multirow}
\usepackage{soul}
\usepackage{titlesec}
\usepackage{pgfplots}
\usepackage{listings}
\usepackage{adjustbox}
\usepackage{setspace,lipsum}
\usepackage{floatrow}
\usepgfplotslibrary{units}

\usepackage{tikz}
\usetikzlibrary{shapes.multipart}
\usepackage{xcolor}

\colorlet{punct}{red!60!black}
\definecolor{background}{HTML}{EEEEEE}
\definecolor{delim}{RGB}{20,105,176}
\colorlet{numb}{magenta!60!black}

\graphicspath{{./Figs/}}

\titleformat{\paragraph}
{\normalfont\normalsize\emph}{\theparagraph}{1em}{}
\titlespacing*{\paragraph}
{0pt}{3.25ex plus 1ex minus .2ex}{1.5ex plus .2ex}

\modulolinenumbers[5]

\usepackage{subcaption}
\usepackage{bm,amsmath,amssymb}
\usepackage{tikz}
    \usetikzlibrary{arrows,positioning,shapes}

\newcommand{\norm}[1]{\left\lVert#1\right\rVert}

{}

\journal{Computational Particle Mechanics}

\begin{document}

\begin{frontmatter}

\title{Lethe-DEM : An open-source parallel discrete element solver with load balancing}

\author[mymainaddress]{Shahab Golshan}
\author[HZG,TUM]{Peter Munch}
\author[Rene]{Rene Gassmöller}
\author[TUM,UU]{Martin Kronbichler}
\author[mymainaddress]{Bruno Blais\corref{mycorrespondingauthor}}
\cortext[mycorrespondingauthor]{Corresponding author}
\ead{bruno.blais@polymtl.com}

\address[mymainaddress]{Research Unit for Industrial Flows Processes (URPEI), Department of Chemical
Engineering, École Polytechique de Montréal, PO Box 6079, Stn Centre-Ville, Montréal,
QC, Canada, H3C 3A7}

\address[HZG]{Continuum Simulations, Institute of Material Systems Modeling, Helmholtz-Zentrum Hereon GmbH, Germany, Max-Planck-Str. 1, 21502 Geesthacht, Germany}

\address[TUM]{Institute for Computational Mechanics, Technical University of Munich, Boltzmannstr. 15, 85748 Garching,
Germany}

\address[Rene]{Department of Geological Sciences, University of Florida, Gainesville, FL, USA}

\address[UU]{Division of Scientific Computing, Department of Information Technology, Uppsala University, Box 337, 751\,05 Uppsala, Sweden}

\begin{abstract}
Approximately $\num{75}\%$ of the raw material and $\num{50}\%$ of the products in the chemical industry are granular materials. The Discrete Element Method (DEM) provides detailed insights of phenomena at particle scale and it is therefore often used for modeling granular materials. However, because DEM tracks the motion and contact of individual particles separately, its computational cost increases non-linearly $O(n_p\log(n_p))$ -- $O(n_p^2)$ depending on the algorithm) with the number of particles ($n_p$). In this article, we introduce a new open-source parallel DEM software with load balancing: Lethe-DEM.
Lethe-DEM, a module of Lethe, consists of solvers for two-dimensional and three-dimensional DEM simulations. Load-balancing allows Lethe-DEM to significantly increase the parallel efficiency by $\approx\num{25}-\num{70}\%$ depending on the granular simulation. We explain the fundamental modules of Lethe-DEM, its software architecture, and the governing equations. Furthermore, we verify Lethe-DEM with several tests including analytical solutions and comparison with other software. Comparisons with experiments in a flat-bottomed silo, wedge-shaped silo, and rotating drum validate Lethe-DEM. We investigate the strong and weak scaling of Lethe-DEM with $\num{1}\leq n_c \leq \num{192}$ and $\num{32}\leq n_c \leq \num{320}$ processes, respectively, with and without load-balancing. The strong-scaling analysis is performed on the wedge-shaped silo and rotating drum simulations, while for the weak-scaling analysis, we use a dam break simulation. The best scalability of Lethe-DEM is obtained in the range of $\num{5000}\leq n_p/n_c \leq \num{15000}$. Finally, we demonstrate that large scale simulations can be carried out with Lethe-DEM  using the simulation of a three-dimensional cylindrical silo with $n_p=\num{4.3}\times10^6$ on \num{320} cores.

\end{abstract}

\begin{keyword}
Discrete Element Methods (DEM),
High-Performance Computing,
Load-balancing,
Silo,
Rotating Drum
\end{keyword}

\end{frontmatter}


\section{Introduction}
Granular material is ubiquitous in nature in the form of soil, sand, or gravel, and the second most used and manipulated material in global industry, right after water \cite{richard2005slow,larsson2019particle}. Approximately half of the products and three quarters of the raw material in the chemical industry is in the form of granular materials \cite{nedderman2005statics,larsson2019particle}. Because of their prominence, an accurate modeling method for granular flows is critical to simulate the efficiency and safety of industrial processes involving such materials. There are two common approaches in the numerical modelling of granular flows, continuum models (Eulerian) and discrete models (Lagrangian). The continuum models, in which a constitutive model describes the relation between stresses and strains, ignore the discrete essence of granular systems and therefore suffer from multiple limitations, for instance in calculating solids holdup \cite{golshan2019modeling}.

The Discrete Element Method (DEM) is a Lagrangian model that simulates the motion and collision of discrete particles with each other and with surfaces \cite{blais2019experimental}. Due to the small modelling scale of individual particles and its discrete nature, DEM is accurate and provides detailed insights of phenomena at particle and system scale \cite{golshan2020review,norouzi2016coupled,blais2019experimental}. DEM has been used to simulate the granular systems in geotechnical \cite{dan2018numerical, coetzee2017calibration}, material processing \cite{wolff2013three,tavarez2007discrete}, mining \cite{cook2002discrete}, chemical \cite{golshan2020review,blais2019experimental}, polymers \cite{ismail2015discrete}, metallurgical \cite{jerier2011study}, pharmaceuticals \cite{ketterhagen2009process}, agricultural \cite{tijskens2003discrete, golshan2019modeling}, and food \cite{boac2014applications} industries. Furthermore, DEM models coupled with Computational Fluid Dynamics (CFD) approaches, a.k.a.~CFD-DEM, provide insight into solid-fluid systems, such as pneumatic conveying, fluidized, and spouted beds \cite{golshan2020review,norouzi2016coupled,blais2016development, berard2020experimental}. In DEM, the flow properties of granular materials are simulated using the interactions between the individual discrete particles. Two main approaches exist to model collisions, namely hard-sphere (event-based) and soft-sphere (time-based) methods \cite{norouzi2016coupled,golshan2020review}. In hard-sphere models, the simulation proceeds collision by collision. As a result, the model is unable to simulate multiple simultaneous collisions or quasi-static systems, making hard-sphere models only suitable for the simulation of dilute systems. On the other hand, in concentrated systems with a high number of particles, the collision frequency is high and multiple simultaneous collisions may occur. Soft-sphere models handle such dense systems by using a small time step ($dt\approx \SI[parse-numbers=false]{10^{-5}-10^{-6}}{\second}$).

\begin{figure} 
	\includegraphics[scale=.4]{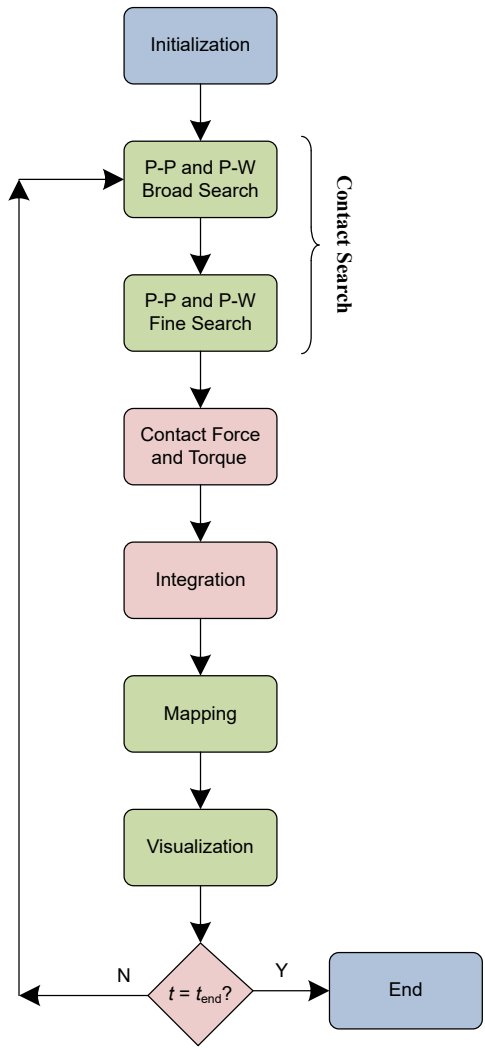}
	\centering
	\caption{Major steps in a conventional DEM simulation. Blue, green and red colors show steps that are performed once, at every n-th time step and at each step throughout the simulation, respectively. $P$ and $W$ stand for particle and wall, respectively.
		\label{fig:Steps}}
\end{figure}

Figure~\ref{fig:Steps} illustrates the major steps of a conventional DEM simulation \cite{golshan2020review}. All these steps have a computational cost that is proportional to the number of particles ($\mathcal{O}(n_p)$), except contact search and collision force calculations. After initialization of the parameters, DEM performs a contact search to obtain a list of all particle-particle and particle-wall contacts. In the worst-case scenario, the DEM has to investigate the possibility of contact of each particle with all other particles and the system walls at each step. This would result in a computational cost of $\mathcal{O}(n_p^2)$ per step. To avoid this, usually a two-step contact search method is employed, often called broad and fine searches, respectively. The broad search detects nearby particles as contact candidates, while the fine search finds the candidate contact pairs in a smaller domain. This two-step contact detection method decreases the computational cost to  $\mathcal{O}(n_p\log{n_p})$. In the next step, the contact list is used to compute the contact forces. Subsequently, the integration step of the DEM updates the accelerations, velocities, and positions of the particles, based on the accumulated forces. According to their new positions, the particles are mapped to the simulation grids for the next time step and visualization purposes.

Several software packages have been developed for DEM simulations, including LIGGGHTS \cite{kloss2012models}, MercuryDPM \cite{weinhart2020fast} and XPS \cite{forgber2020extended}. In this work, we introduce a new open-source, parallel and load-balanced software for the simulation of granular systems: Lethe-DEM. Lethe-DEM is a DEM solver in the framework of Lethe \cite{blais2020lethe}, a high-order CFD solver to simulate incompressible flows and targeting practice-oriented chemical engineering applications. Lethe is built upon deal.II \cite{arndt2020deal}, a well-established open-source finite element modelling library. Lethe-DEM uses the background mesh infrastructure of the deal.II library, which allows it to reuse many functionalities that have been implemented with FEM in mind and provides several advantages. First of all, it enables the use of high-order finite element mappings, enabling the simulation of complex geometries (such as cylinders and spheres) with high accuracy. Additionally, Lethe-DEM inherits deal.II's high-performance infrastructure, such as SIMD vectorization \cite{kronbichler2012} and, more importantly, parallelization capabilities with the message passing interface (MPI) including load-balancing capabilities through the p4est library \cite{burstedde2011p4est, burstedde2020}. Using load-balancing, Lethe-DEM redistributes the computational load among the CPUs throughout a simulation by ensuring that all cores have a similar amount of particles. Load-balancing diminishes the possibility of having idle cores during a simulation and decreases the simulation time significantly, especially in large-scale systems or when complex and sparse geometries are considered. Due to this redistribution of the computational load via load-balancing, Lethe-DEM shows good strong scaling.

The remainder of this work is organized as follows. First, we present the governing equations of the DEM models in Lethe-DEM in Section~\ref{section:equations}. Then, Section \ref{section:software} describes the architecture of Lethe-DEM, including contact detection, parallelization and load-balancing strategies. In Section \ref{section:verification} we verify and validate the software with available correlations, results of other DEM codes and experimental data. Additionally, we investigate both the strong and weak scaling capabilities of the software. Finally, Section \ref{section:conclusions} concludes the article and presents the future works of the software.

\section{Governing equations and solution strategies}\label{section:equations}
\begin{figure} 
	\includegraphics[scale=0.65]{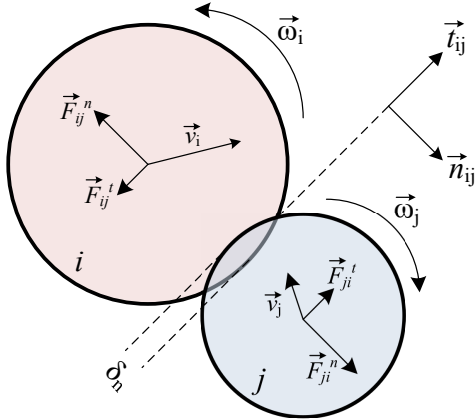}
	\centering
	\caption{Typical particle-particle contact in the framework of soft-sphere DEM. For a description of the shown quantities see equations~\eqref{eq:main1} and~\eqref{eq:main2}.
		\label{fig:Contact}}
\end{figure}

In DEM, Newton's second law of motion describes the evolution of the position and the velocity of all the particles through time:
\begin{equation}\label{eq:main1}
	m_i\bm{a}=m_i\frac{d\bm{v_i}}{dt}=\sum_{j\in \mathcal C_i} (\bm{F}_{ij}^n + \bm{F}_{ij}^t) + m_i\bm{g} + \bm{F}_i^\text{ext}
\end{equation}
\begin{equation}\label{eq:main2}
	I_i\frac{d\bm{\omega_i}}{dt}=\sum_{j\in \mathcal C_i} (\bm{M}_{ij}^t + \bm{M}_{ij}^r) +  \bm{M}_i^\text{ext}
\end{equation}
where $i$ is the particle index, $m_i$ is the mass of particle $i$, $\bm{v}_i$ the velocity of particle $i$, $\bm{a}_i$ the acceleration of particle $i$, $t$ denotes time, $j$ runs through all particles $\mathcal C_i$ in the contact list of particle $i$, $\bm{F}_{ij}^n$ and $\bm{F}_{ij}^t$ are normal and tangential contact forces due to the contact between particles $i$ and $j$, $\bm{g}$ is the gravitational acceleration, $\bm{F}^\text{ext}$ encompasses all other external forces, $I_i$ is the moment of inertia of particle $i$, $\bm{\omega_i}$ denotes the angular velocity of particle $i$, $\bm{M}_{ij}^t$ and $\bm{M}_{ij}^r$ are tangential and rolling friction torques due to the contact between particle $i$ and $j$, and $\bm{M}^\text{ext}$ denotes all other external torques. Figure~\ref{fig:Contact} shows a typical contact between two particles in the framework of soft-sphere DEM. This figure illustrates some of the variables introduced in equations \eqref{eq:main1} and \eqref{eq:main2}.

\subsection{Contact force and torque models}
The particles can overlap in the soft-sphere model (see Figure \ref{fig:Contact}), and the contact forces in the normal and tangential directions are defined using the normal and tangential overlaps. The normal overlap ($\delta_n$) between the particles $i$ and $j$ is computed as:

\begin{equation}\label{eq:deltan}
	\delta_n=R_i+R_j-\norm{\bm{x}_{j}-\bm{x}_{i}}
\end{equation}
where $\bm{x}_i$ and $\bm{x}_j$ are the position vectors of the particle centers, and $R_i, R_j$ are radii of particles $i$ and $j$, respectively. Consequently, the normal overlap is positive when the distance between two particles $i$ and $j$ is smaller than the sum of their radii, that is, when particles are undergoing collision. 

The contact normal vector ($\bm{n}_{ij}$), as illustrated in Figure~\ref{fig:Contact}, is a vector pointing from particle $i$ to particle $j$ in a collision, computed as:

\begin{equation}\label{eq:vecn}
	\bm{n}_{ij}=\frac{\bm{x}_{j}-\bm{x}_{i}}{\left|\bm{x}_{j}-\bm{x}_{i}\right|}
\end{equation}

The tangential overlap ($\bm{\delta}_t$) in the contact models of Lethe-DEM depends on the contact history. At the beginning of each contact, the tangential overlap is equal to zero. The following equation is then used to update the tangential overlap during a contact \cite{norouzi2016coupled}:
\begin{equation}\label{eq:delta_t}
	\bm{\delta}_{ij}^{t,\text{new}}=\bm{\delta}_{ij}^{t,\text{old}}+\bm{v}_{rt}dt
\end{equation}

Here, $v_{rt}$ is the contact relative velocity in tangential direction and is computed as described below.

This approach is different from the one taken in MercuryDPM \cite{weinhart2020fast}, which uses a two-step procedure to update the tangential overlap during a contact. This two-step procedure consists of removing the normal component of the tangential overlap using the updated normal vector in the first step, and updating the tangential overlap using the updated relative velocity in the tangential direction in the second step. Even though this could theoretically lead to some minor differences when a particle rolls on top of another particle, we did not observe any significant benefit in several benchmarks. As a result, we followed the procedure which is also utilized in other software packages including LIGGGHTS \cite{kloss2012models}. The contact relative velocity in the tangential direction ($\bm{v}_{rt}$) is given by:
\begin{equation}\label{eq:vrt}
	\bm{v}_{rt}=\bm{v}_{ij}-\bm{v}_{rn}
\end{equation}
where the contact relative velocity ($\bm{v}_{ij}$) and relative velocity in the normal direction ($\bm{v}_{rn}$) are given respectively given by:
\begin{equation}
	\bm{v}_{ij}=\bm{v}_i-\bm{v}_j+\left(R_i\bm{\omega}_i+R_j\bm{\omega}_j\right)\times\bm{n}_{ij}
\end{equation}
\begin{equation}
	\bm{v}_{rn}=\left(\bm{v}_{ij}.\bm{n}_{ij}\right)\bm{n}_{ij}
\end{equation}

The computed normal and tangential overlaps are then used to calculate normal and tangential contact forces. For the calculation of contact forces and torques, previous studies have proposed several models. Linear viscoelastic models (a.k.a. Hookean models) \cite{cundall1979discrete}, in which the contact force is a linear function of the contact overlap, and non-linear viscoelastic models (a.k.a. Hertzian models), in which the relation between contact force and overlap is non-linear, are the most well-known models. 

In linear and non-linear viscoelastic models, normal and tangential contact forces are calculated using the following equations \cite{cundall1979discrete, tsuji1992lagrangian}:
\begin{equation}\label{eq:normal_force}
	\bm{F}_{ij}^n=-(k_n\delta_n)\bm{n}_{ij}-(\eta_n\bm{v}_{rn})
\end{equation}
\begin{equation}\label{eq:tangential_force}
	\bm{F}_{ij}^t=-(k_t\bm{\delta}_t)-(\eta_t\bm{v}_{rt})
\end{equation}
where $k$ is the spring constant and $\eta$ is the damping constant. In linear models, spring and damping constants do not depend on the normal or tangential overlaps, whereas in non-linear models, these constants are a function of the overlaps. 

Several models have been proposed to calculate the spring and damping constant in the context of linear models. In Lethe-DEM, we use the model expressed with the following equation \cite{kloss2012models}:
\begin{equation}\label{eq:linearkn}
	k_n=\frac{16}{15}\sqrt{R_{e}}Y_{e}\left(\frac{15m_{e}V^2}{16\sqrt{R_{e}}Y_{e}}\right)^{0.2}
\end{equation}
\begin{equation}\label{eq:linearetan}
	\eta_n=\sqrt{\frac{4m_{e}k_n}{1+\left(\frac{\pi}{\ln{e}}\right)^2}}
\end{equation}
where $R_{e}, Y_{e}, m_{e}, V$ and $e$ are the effective radius, effective Young's modulus, effective mass, characteristic impact velocity, and coefficient of restitution, respectively. Spring and damping constants in the tangential direction are also calculated using Equations~\eqref{eq:linearkn} and \eqref{eq:linearetan}.

On the other hand in non-linear viscoelastic models, the following equations calculate the normal and tangential spring and damping constants \cite{kloss2012models}:
\begin{subequations}
\begin{equation}\label{eq:nonlinearkn}
	k_n=\frac{4}{3}Y_{e}\sqrt{R_{e}\delta_n}
\end{equation}
\begin{equation}\label{eq:nonlinearetan}
	\eta_n=-2\sqrt{\frac{5}{6}}\beta\sqrt{S_nm_{e}}
\end{equation}
\begin{equation}\label{eq:nonlinearkt}
	k_t=8G_{e}\sqrt{R_{e}\delta_n}
\end{equation}
\begin{equation}\label{eq:nonlinearetat}
	\eta_t=-2\sqrt{\frac{5}{6}}\beta\sqrt{S_tm_{e}}
\end{equation}
\end{subequations}

Table~\ref{table:equation_parameters} reports the parameters of Equations\eqref{eq:nonlinearkn}--\eqref{eq:nonlinearetat}.

\renewcommand{\arraystretch}{1.5} 
\begin{table}[!htbp]
\centering
	\begin{tabular}{ |c|c|c|c| } 
		\hline
		\textbf{Parameter} & \textbf{Equation} \\
		\hline
		Effective mass & $\frac{1}{m_{e}}=\frac{1}{m_i}+\frac{1}{m_j}$ \\ 
		Effective radius & $\frac{1}{R_{e}}=\frac{1}{R_i}+\frac{1}{R_j}$ \\ 
		Effective shear modulus & $\frac{1}{G_{e}}=\frac{2(2-\nu_i)(1+\nu_i)}{Y_i}+\frac{2(2-\nu_j)(1+\nu_j)}{Y_j}$ \\ 
		 Effective Young's modulus & $	\frac{1}{Y_{e}}=\frac{\left(1-\nu_i^2\right)}{Y_i}+\frac{\left(1-\nu_j^2\right)}{Y_j}$ \\ 
		 $\beta$ & $\beta=\frac{\ln{e}}{\sqrt{\ln^2{e}+\pi^2}}$ \\ 
		 $S_n$ & $S_n=2Y_{e}\sqrt{R_{e}\delta_n}$ \\ 
		 $S_t$ & $S_t=8G_{e}\sqrt{R_{e}\delta_n}$ \\ 
		\hline
	\end{tabular}
\caption{Parameters in Equations~\eqref{eq:nonlinearkn}-\eqref{eq:nonlinearetat}}
\label{table:equation_parameters}
\end{table}

The following equation calculates the tangential torque in Lethe-DEM:
\begin{equation}\label{eq:torque_t}
	\bm{M}_{ij}^t=R_i\bm{n}_{ij}\times\bm{F}_{ij}^c
\end{equation}

Two models, namely constant \cite{zhou1999rolling, norouzi2016coupled} and viscous \cite{brilliantov1998rolling, norouzi2016coupled}, are available for the calculation of the rolling friction torque. Table \ref{table:rolling_torque} reports these two models and their corresponding parameters.

\begingroup
\setlength{\tabcolsep}{10pt} 
\renewcommand{\arraystretch}{1.5} 
\begin{table}[!htbp]
\centering
	\begin{tabular}{ |c|c|c|c| } 
		\hline
		\textbf{Model} & \textbf{Equation} \\
		\hline
		Constant torque model \cite{zhou1999rolling} & $\bm{M}_{ij}^r=-\mu_rR_{e}\left|\bm{F}_{ij}^n\right|\bm{\hat{\omega}}_{ij}$ \\ 
		Viscous torque model \cite{brilliantov1998rolling} & $\bm{M}_{ij}^r=-\mu_rR_{e}\left|\bm{F}_{ij}^n\right|\left|\bm{V}_{\omega}\right|\bm{\hat{\omega}}_{ij}$ \\ 
		Parameters & $\bm{\hat{\omega}}_{ij}=\frac{\bm{\omega}_i-\bm{\omega}_j}{\left|\bm{\omega}_i-\bm{\omega}_j\right|}$ \\ 
		    & $\bm{V}_{\omega}=\left(\bm{\omega}_i\times R_i\bm{n}_{ij}-\bm{\omega}_j\times R_j\bm{n}_{ji}\right)$ \\ 
		\hline
	\end{tabular}
\caption{Rolling friction torque models in Lethe-DEM and the corresponding parameters.  $\bm{F}_{ij}^n$ and $\mu_r$ are contact force in normal direction and rolling friction coefficient, respectively.}
\label{table:rolling_torque}
\end{table}
\endgroup

If $\left|\bm{F}_{ij}^t\right|\geq\mu\left|\bm{F}_{ij}^n\right|$ occurs during a collision, the Coulomb's criterion is violated (a.k.a. gross sliding). In this case, the magnitude of the exerted kinetic friction is independent of the magnitude of particles' velocity, and the tangential overlap is limited to \cite{cundall1979discrete}:
\begin{equation}\label{eq:coulomb_overlap}
\bm{\delta}_t=\frac{\tilde{\bm{F}}_{ij}^t}{-k_t}
\end{equation}
where $\tilde{\bm{F}}_{ij}^t$ is the limited elastic tangential force (tangential force without the damping force),
\begin{equation}\label{eq:elastic}
\tilde{\bm{F}}_{ij}^t=\hat{\bm{F}}_{ij}^t+\eta_t\bm{v}_{rt}
\end{equation}
in which $\hat{\bm{F}}_{ij}^t$ is the limited tangential force:
\begin{equation}\label{eq:coulomb_overlap3}
\hat{\bm{F}}_{ij}^t=\mu\left|\bm{F}_{ij}^n\right|\frac{\bm{F}_{ij}^t}{\left|\bm{F}_{ij}^t\right|}
\end{equation}

To limit the tangential force to the Coulomb's limit, two different approaches are common in the literature\cite{weinhart2020fast,tsuji1992lagrangian,kloss2012models, norouzi2016coupled}. Some researchers only limit the tangential force to the Coulomb's limit, while others limit the tangential overlap first and then recalculate the tangential force using this limited overlap. In Lethe-DEM, we use the second approach, similar to the method used in LIGGGHTS \cite{kloss2012models}, to ensure that the tangential overlap does not accumulate during gross sliding. Furthermore, we noticed that using the equation $\bm{\delta}_t={\bm{F}_{ij}^t}/{k_t}$ is not consistent with tangential force models including damping force. Tangential force models with damping lose their continuity when they exceed the Coulomb's limit (with $\bm{\delta}_t={\bm{F}_{ij}^t}/{k_t}$). As a result, we use  equations~\eqref{eq:coulomb_overlap}--\eqref{eq:coulomb_overlap3} to maintain the continuity of the tangential force when it is limited to the Coulomb's limit.

For particle-wall contacts, we apply the same model and equations of particle-particle contacts. We use the background triangulation and mapping, which are available using deal.II, to obtain the normal vectors at the location of the contact and the normal overlap with each boundary wall. 

\subsection{Integration methods}
Equation~\eqref{eq:main1} converts the net exerted force on a particle to its acceleration at each DEM time step. Several integration methods can be used in DEM \cite{norouzi2016coupled}. In Lethe-DEM, we have implemented the Explicit Euler and the Velocity Verlet methods.

\subsubsection{Explicit Euler}
In the explicit Euler method, the following equations calculate the velocity and position of a particle at each time step:
\begin{equation}\label{eq:integ_EE}
	\bm{v}_i^{n+1}=\bm{v}_i^{n}+\bm{a}_i^{n}dt
\end{equation}
\begin{equation}
	\bm{x}_i^{n+1}=\bm{x}_i^{n}+\bm{v}_i^{n}dt
\end{equation}
in which $n$ and $n+1$ denote the current and the next step, respectively. This scheme is first order accurate for both position and velocity. It is very simple to implement, but its accuracy is limited \cite{norouzi2016coupled}. The same method is applied to the calculation of angular velocity using the momentum exerted on a particle.

\subsubsection{Velocity Verlet}
The velocity Verlet integration scheme uses a half-step velocity $\bm{v}_i^{n+\frac{1}{2}}$ to update the position of the particles:
\begin{equation}\label{eq:halfstep}
	\bm{v}_i^{n+\frac{1}{2}}=\bm{v}_i^{n}+\bm{a}_i^{n}\frac{dt}{2}
\end{equation}
\begin{equation}
	\bm{x}_i^{n+1}=\bm{x}_i^{n}+\bm{v}_i^{n+\frac{1}{2}}dt
\end{equation}

The new velocity of the particle is:
\begin{equation}\label{eq:newvv}
	\bm{v}_i^{n+1}=\bm{v}_i^{n+\frac{1}{2}}+\bm{a}_i^{n+1}\frac{dt}{2}
\end{equation}

When the force only depends on the position, the Velocity Verlet scheme is second-order accurate for position and velocity. In this context, the Velocity Verlet scheme is symplectic (see \cite{delacroix2020simulation} for a full demonstration). When the force depends on the velocity, as is the case when the coefficient of restitution is not one, the scheme is first-order accurate for both position and velocity \cite{delacroix2020simulation}. In practice, however, it is still significantly more accurate than an explicit Euler scheme. Since the only velocity-dependent force in DEM are dissipative forces, the loss of symplectic characteristic of the integration scheme does not pose a problem since there is no more energy to conserve.

\section{Software description and architecture}\label{section:software}
In this section, the architecture of Lethe-DEM as well as the main features and algorithms of this software are explained.

\subsection{Overview of Lethe-DEM}
\begin{figure} 
	\includegraphics[scale=.5]{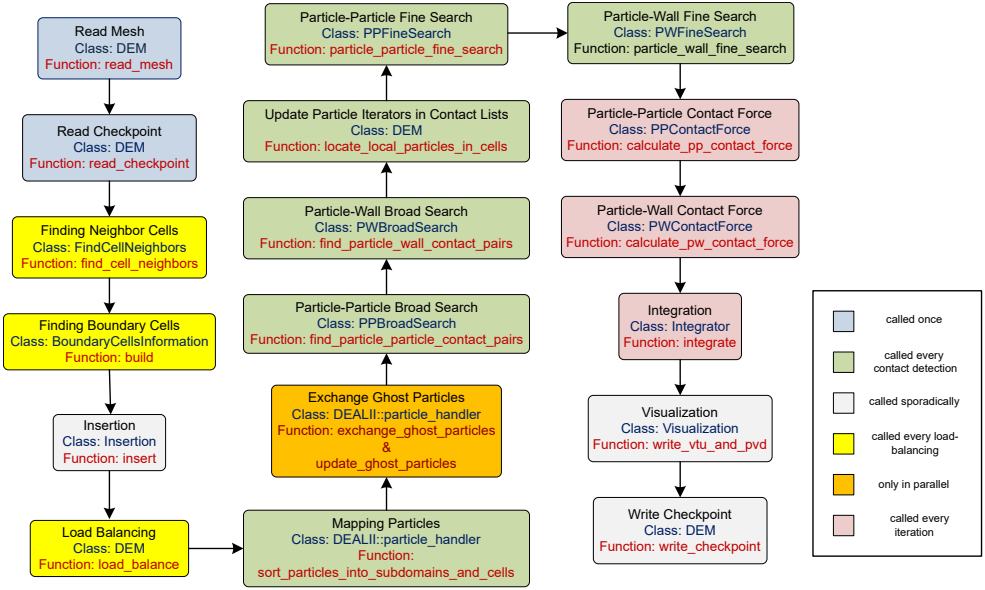}
	\centering
	\caption{Lethe-DEM fundamental steps, classes and function. 
		\label{fig:LetheAlgorithm}}
\end{figure}

The fundamental steps, features, and architecture of Lethe-DEM are illustrated in Figure~\ref{fig:LetheAlgorithm}. The main steps are divided into six main groups depending if the step is called once, at every contact detection, sporadically, at every load balancing operation, only in parallel or at every iteration.

At the beginning, the simulation domain is defined using a triangulation (grid), which defines the system outer and inner boundaries. Similar to Lethe \cite{blais2020lethe}, unstructured meshes (using
the GMSH \cite{geuzaine2009gmsh} file format) and native meshes produced by the deal.II library (ranging from a simple ball and cube to more complex ones
like airfoils) are supported. Lethe-DEM supports writing and reading checkpoint files, which enables the user to quickly restart an unfinished simulation.

\begin{figure} 
	\includegraphics[scale=.5]{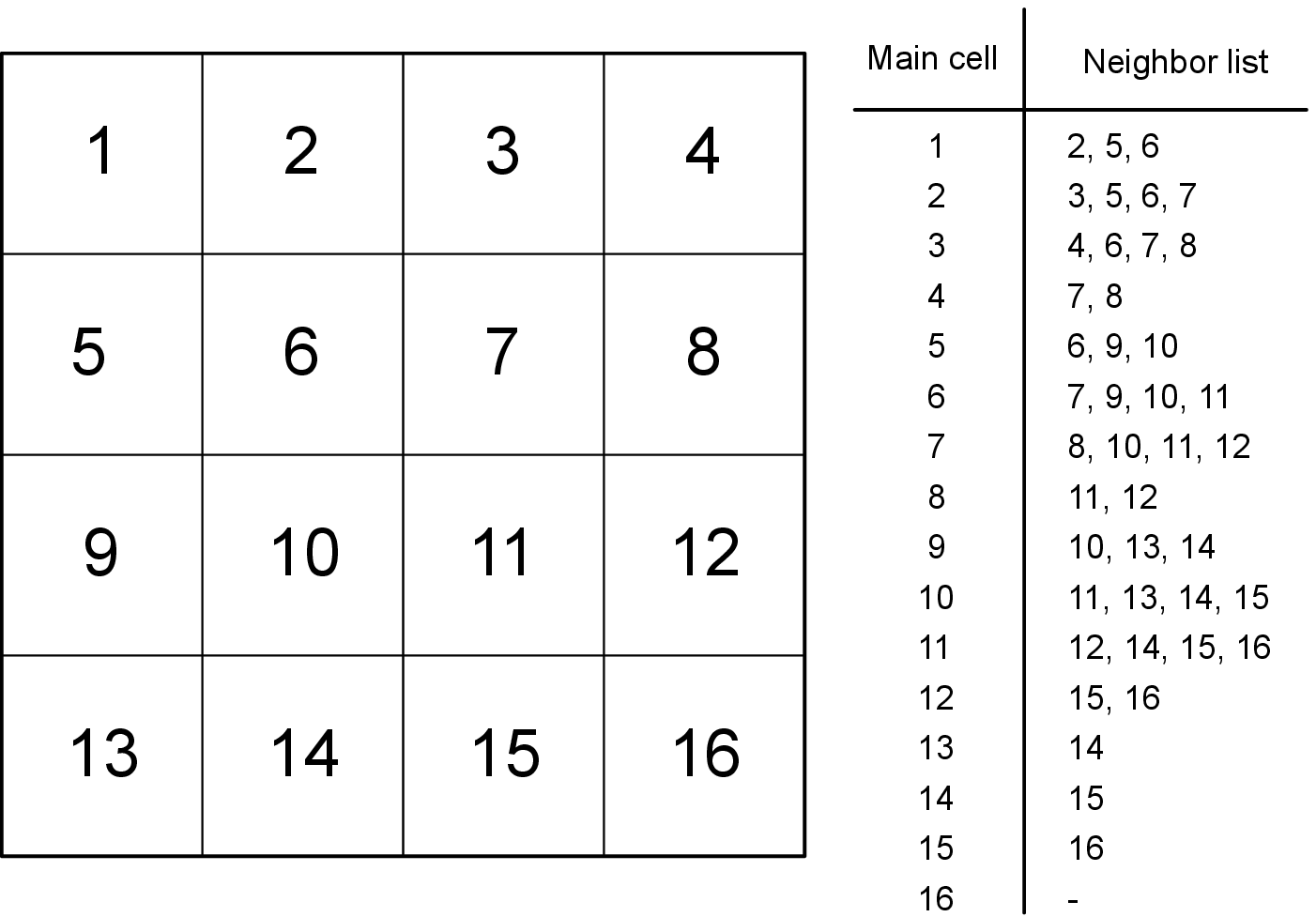}
	\centering
	\caption{A sample triangulation and the corresponding cell neighbors lists. 
		\label{fig:SampleNeighbor}}
\end{figure}

 Once the triangulation has been initialized, the neighbor lists of all cells  are produced (\texttt{Finding Neighbor Cells} block in Figure~\ref{fig:LetheAlgorithm}). Cell neighbor lists is a data container in which the neighbors of all the cells sharing a vertex/node are stored. To make broad and fine searches more efficient, repetitions are avoided in the neighbor list. For instance, if cell $j$ is in the neighbor list of cell $i$, we do not reconsider cell $i$ in the neighbors list of cell $j$. Figure~\ref{fig:SampleNeighbor} gives as an example a two-dimensional triangulation and its cell neighbor lists. The same procedure applies for obtaining neighbor lists in three-dimensional triangulations. In load-balancing iterations, cell neighbor lists have to be reconstructed since the owners of the cells change. The details of load balancing are explained in more detail in Section \ref{sec::parallelization}. Similar to the cell neighbor lists, boundary cells are also identified within the triangulation (\texttt{Finding Boundary Cells} block in Figure~\ref{fig:LetheAlgorithm}). For the sample triangulation in Figure~\ref{fig:SampleNeighbor}, cells 1, 2, 3, 4, 5, 8, 9, 12, 13, 14, 15 and 16 are the boundary cells.

In the insertion step (\texttt{Insertion} block in Figure~\ref{fig:LetheAlgorithm}), particles with specified properties (e.g., size) are inserted in the simulation domain. In Lethe-DEM, two particle size types, namely uniform and normal size distributions, can be used. A uniform size distribution results in a constant diameter for all particles. A normal size distribution for a parameter $\xi$ is defined as:
\begin{equation}
	f(\xi)=\frac{1}{\sigma\sqrt{2\pi}}e^{\frac{1}{2}\left(\frac{\xi-\mu}{\sigma}\right)^2}
\end{equation}
where $\mu$ and $\sigma$ are average and standard deviation of the normal distribution. Multiple particle types, with different size distributions (uniform or normal) and mechanical properties can be inserted by defining a new particle type in the parameter handler file. Given a batch of particles, the software searches the surrounding cell of each particle and assigns the responsibility of the particle to that cell. Using this information, broad and fine contact searches obtain particle-particle (\texttt{Particle-Particle Broad Search} and \texttt{Particle-Particle Fine Search} blocks in Figure~\ref{fig:LetheAlgorithm}) and particle-wall (\texttt{Particle-Wall Broad Search} and \texttt{Particle-Wall Fine Search} blocks in Figure~\ref{fig:LetheAlgorithm}) contact lists. Lethe-DEM does not call broad and fine contact searches every iteration. It supports two approaches to find contact search iterations: \texttt{constant} and \texttt{dynamic}. In the \texttt{constant} approach, the simulation calls broad and fine searches at a frequency of every $n$ time steps, defined by a user defined parameter. On the other hand, in the \texttt{dynamic} method, Lethe-DEM automatically estimates the iterations for calling broad and fine searches by tracking the maximal displacement of particles and comparing it with a user-defined value.

In the next step, Equations \eqref{eq:deltan}--\eqref{eq:normal_force} calculate the contact forces of the pairs present in the particle-particle and particle-wall contact lists (\texttt{Particle-Particle Contact Force} and \texttt{Particle-Wall Contact Force} blocks in Figure~\ref{fig:LetheAlgorithm}). After calculating the contact forces, the integration class (\texttt{Integration} block in Figure~\ref{fig:LetheAlgorithm}) updates the locations and velocities of all the particles by using Equations \eqref{eq:integ_EE}--\eqref{eq:newvv}. Finally, when required, the software outputs the necessary post-processing files (\texttt{Visualization} block in Figure~\ref{fig:LetheAlgorithm}) and the checkpoints (\texttt{Write Checkpoint} block in Figure~\ref{fig:LetheAlgorithm}).

\subsection{Contact detection algorithm}
\begin{figure*} 
	\includegraphics[scale=.6]{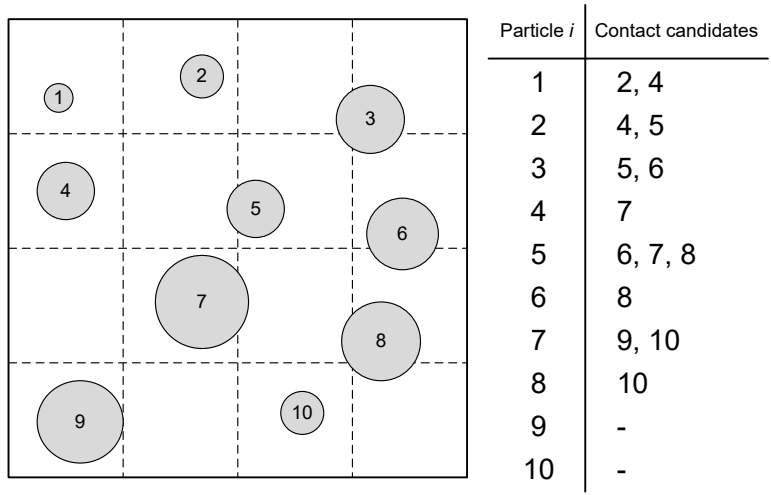}
	\centering
	\caption{Contact pair candidates (which is the output of broad search and input of fine search) in a sample system. 
		\label{fig:ContactPair}}
\end{figure*}

In the broad search, we use the cell neighbor lists to find contact pair candidates. To this end, Lethe-DEM loops through all the cells in the triangulation: for each cell it loops through all the particles, and for each particle, it adds other particles in the main cell or in neighbor cells to the contact pair candidates. Figure~\ref{fig:ContactPair} shows the contact pair candidates in a sample system. 

\begin{figure} 
	\includegraphics[scale=1]{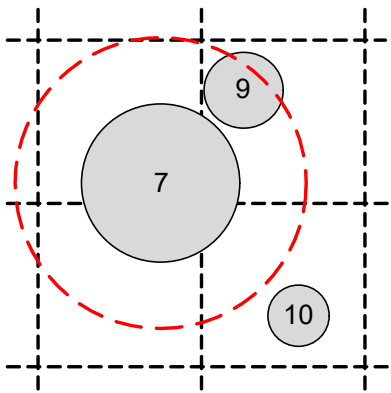}
	\centering
	\caption{A sample system for showing contact list (which is the output of fine search and input of contact force class) members. 
		\label{fig:FineSearch}}
\end{figure}

The fine search uses these contact pair candidates as an input, and calculates the distance between these pair candidates. If the distance between a pair is smaller than a used-defined threshold (generally $1.3 d_p$, where $d_p$ is the maximum particle diameter), the fine search adds the pair to the contact list. For instance, in the system illustrated in Figure~\ref{fig:FineSearch}, the red circle shows the domain for the fine search. In this system, particle 9 is in the contact list of particle 7, while particle 10 is not in this list. This contact list is the output of the fine search. The contact forces are calculated using this contact list. Contact list as well as contact pair candidates are updated at the contact search iterations.

\subsection{User input and parameter files}
To enable full control over the simulations, Lethe-DEM is parameterized using an input file in the native deal.II \textit{prm} format. There are separated sections, such as \texttt{Simulation control}, \texttt{Model parameters}, \texttt{Physical properties}, \texttt{In\-sertion information}, \texttt{Mesh reading/generation}, and \texttt{Boundary motion} in an input file. Listing~\ref{lst::prm} in Appendix A shows examples of an input parameter file for a DEM simulation.

\subsection{Version control, continuous integration, and testing}
Lethe-DEM is directly integrated within Lethe~\cite{blais2020lethe}. It is built as separate executable and is configured using CMake \cite{martin2010mastering}. A public GitHub repository \url{https://github.com/lethe-cfd/lethe} keeps Lethe under version control. The Wiki documentation of the project is also on the GitHub repository. Lethe-DEM is distributed with several unit tests for individual functions and classes, and application tests for all the solvers. GitHub Actions instances are used for continuous integration and to verify all the tests (including unit and application tests). This ensures the stability of the master branch and ensure quality control when merging pull requests. Lethe is distributed under an LGPL 3.0 license.

\subsection{Parallelization}\label{sec::parallelization}
\begin{figure} 
	\includegraphics[scale=.4]{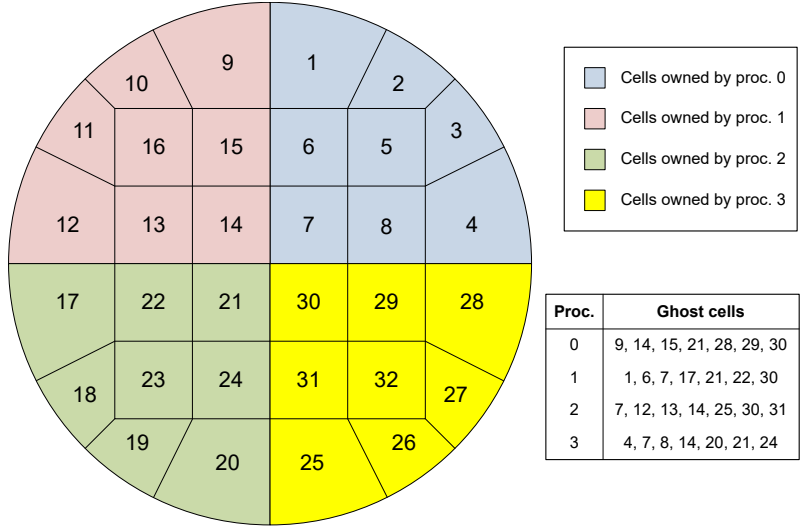}
	\centering
	\caption{A sample distributed triangulation on four processes, local and ghost cells are specified for each process. 
		\label{fig:ParallelCells}}
\end{figure}

Lethe-DEM is parallelized using MPI \cite{gropp1996high}. This allows the simulation of large problems on
distributed computer architectures. Lethe-DEM leverages the p4est library via deal.II to handle mesh partitioning \cite{burstedde2011p4est, bangerth2012}. First, a coarse triangulation is created, either from a GMSH file or a deal.II native triangulation. This coarse triangulation, which in most cases consists of not more than 100--10,000 cells, is replicated on all processes. This triangulation is then refined locally or globally to a desired element size in a forest-of-tree manner, partitioning the final triangulation with a well-established domain-decomposition approach. Each locally-owned subdomain is surrounded by a halo layer of the width of a single cell. Particles which reside in the ghost cells of a process are defined as ghost particles. In Figure~\ref{fig:ParallelCells}, a sample domain is distributed among four processes. For each process, we specify the locally owned and the ghost cells.

All the operations in Lethe-DEM except particle-particle and particle-wall contacts occur in the local domain of each process. For each process, we categorize the particle-particle collisions into two types:  local-local and local-ghost contacts. In local-local contacts between particles $i$ and $j$, the process owns both particles $i$ and $j$, while in local-ghost contacts, the process owns particle $i$, while another process owns particle $j$ located in a ghost cell. Therefore, the processes fully handle local-local collisions, while for local-ghost collisions, only the exerted contact force on particle $i$ and the associated integration is performed using the information (position, velocity, etc.) of the ghost particle of index $j$. The \texttt{exchange\_ghost\_particles} and  \texttt{update\_ghost\_particles} functions update the information of location and properties of the ghost particles.

The \texttt{exchange\_ghost\_particles} function creates and updates the ghost particles by removing particles that have left the ghost cells and as a consequence have become unnecessary, as well as by creating the required ghost particles incoming from other processes. Furthermore, it sets up the communication patterns while the \texttt{update\_ghost\_particles} function only updates the properties and location of the ghost particles. Therefore, we call the former only at contact detection iterations, where we are locating particles in the cells by calling the \texttt{sort\_particles\_into\_subdomains\_and\_cells} function. In other iterations, we call the significantly cheaper \texttt{update\_ghost\_particles} function.

\subsection{Load balancing}
Lethe-DEM supports load-balancing for parallel computations by using p4est's and deal.II's functionalities. Each cell owned by the local process is assigned a weight according to its computational load. Lethe-DEM then distributes the sum of these weights evenly among the available processes. Users can specify the load-balancing frequency in the parameter-handler file. At the load-balancing iterations, Lethe-DEM re-equalizes the computational loads on the process according to the number of particles in the distributed simulation domain. Here, a load $L_c$ is attributed to each process, which is a linear combination of the number of cells $n_e$ and particles $n_p$ owned by the process:
\begin{align}
    L_{c}=\alpha n_p+\beta n_e
\end{align}
where $\alpha$ and $\beta$ are weights of particles and cells. During the redistribution, this load is balanced between the processes. Currently, the defined ratio of the weights is $\alpha/\beta=10$. Considering the high number of particles compared to the number of cells in a DEM simulation, this weighting strategy focuses mainly on the number of particles owned by each process. We should mention that future investigations are required to optimize the ratio of the weights and investigate the performance of this linear weighting strategy, in particular once we tackle the coupling of DEM and CFD which possibly have conflicting load balancing needs. The same load balancing algorithm applied to a different application has been shown to only modestly influence the performance as long as the ratio remains reasonably close to an optimal ratio that is model dependent  \cite{gassmoller2018flexible}.

\subsection{Other features}
Some of the other useful features of Lethe-DEM are covered in this section. First of all, Lethe-DEM makes use of C++ templates to create two separate solvers for two-dimensional (\texttt{dem\_2d}), and three-dimensional (\texttt{dem\_3d}) simulations. Hence, both 2D and 3D simulations are supported without loss of performance using the same code base. Another important feature of Lethe-DEM is particle-line and particle-point contacts. Lethe-DEM automatically checks the input triangulation for possible boundary lines (edges) or points (vertices). Particles may collide with boundary faces, as well as boundary lines and points in specific geometries. If boundary lines or points exist in the input triangulation, the software considers the contact between particles and these boundary elements. Boundary lines may only exist in three-dimensional triangulations, while boundary points may exist in both two-dimensional and three-dimensional triangulations.

\begin{figure} 
	\includegraphics[scale=.3]{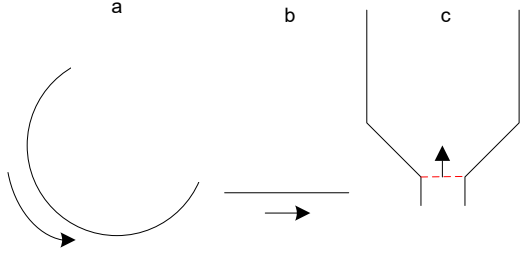}
	\centering
	\caption{Schematics of a: rotational boundary motion, b: translational boundary motion, and c: a floating wall (red dashed line) with its normal vector (arrow).
		\label{fig:boundary_motion}}
\end{figure}

Another feature of Lethe-DEM is the motion of boundaries. Boundaries can move in Lethe-DEM. Currently, users can define two types of motion, namely rotational (Figure~\ref{fig:boundary_motion}a) and translational (Figure~\ref{fig:boundary_motion}b), in the parameter handler file. By specifying the boundary id, type of motion, and motion properties, Lethe-DEM implements the motion in the simulation. Another feature of Lethe-DEM is the definition of floating walls, as illustrated in Figure~\ref{fig:boundary_motion}c. By specifying the normal vector of the floating wall (arrow in Figure~\ref{fig:boundary_motion}c), a point on the wall, start and end times of the wall, Lethe-DEM defines it automatically in the simulation. This feature is especially convenient when simulating filling and discharge of particles in containers.

\section{Verification and validation}\label{section:verification}
All the major functions and modules of Lethe are tested and verified using unit tests. New features added to the software are tested with at least one test to guarantee the accuracy and reproducibility of its outputs. Besides these tests, we have compared the outputs of Lethe-DEM with experimental data and simulations made using other software. We present the result of the comparisons in the following.

\subsection{Verification}
In this section, we compare the outputs of Lethe-DEM and analytical solutions by three test cases.

\subsubsection{Particle trajectory before and after a collision}
In this test, we compare the height of a particle before and after a collision with a wall. The accuracy of particle-wall collisions extends to particle-particle collisions, since the same model and calculation procedure are employed for both. 
In Figure \ref{fig:particle_pos}, we show the results of this comparison. The analytical solution for the position of the particle consists of three parts: free fall of the particle before the collision, contact with the wall, and motion after the collision. Equations~\eqref{eq:position1}--\eqref{eq:position3} express the location of the particle during these steps \cite{garg2012open,norouzi2017new}.

\begin{subequations}
\begin{equation}\label{eq:position1}
y(t)=h_0-\frac{gt^2}{2}
\end{equation}

\small
\begin{align}\label{eq:position2}
y(t-t_1)=&\left(\alpha_1\cos(w(t-t_1))+\alpha_2\sin(w(t-t_1))\right)\exp(-\beta_1 \omega_0(t-t_1)) + r_p-\alpha_1
\end{align}

\begin{equation}\label{eq:position3}
y(t-t_2)=r_p+v_2(t-t_2)-0.5g{(t-t_2)}^2
\end{equation}
\end{subequations}

where 
\begin{align}
  \alpha_1=\frac{g}{\omega_0^2} \\
  \alpha_2=\frac{-\sqrt{2g(h_0-r_p)}+\frac{\beta g}{\omega_0}}{w} \\
  \beta_1=\frac{\eta_n}{2\sqrt{k_nm_i}}\\
  \omega_0=\sqrt{\frac{k_n}{m_i}} \\
  w=\sqrt{1-\beta^2\omega_0}
\end{align} and $h_0$, $y$, $t_1$, $t_2$ and $v_2$ are initial particle height, vertical position, end times of steps one and two, and velocity of the particle at the end of step two, respectively. Average absolute error (AAE) values (0.82\%, 0.68\% and 0.32\% for coefficients of restitution $e=$ 0.5, 0.7 and 0.9, respectively for $\Delta t=\SI[parse-numbers=false]{7\times10^{-6}}{\second}$) demonstrate the accuracy of Lethe-DEM in modelling the contacts. AAE is defined as:
\begin{equation}\label{eq:AAE}
\text{AAE}=\frac{\sum_{i=1}^{l} \left|\frac{f^\text{exp}-f^\text{sim}}{f^\text{exp}}\right|}{l} \cdot 100\%
\end{equation}
 where $l$, $f^\text{exp}$ and $f^\text{sim}$ denote the number of data points, the expected value from the analytical solution, and the simulated value from Lethe-DEM.

\input{Figs/Figure_particle_pos}

\subsubsection{Orders of convergence of the integration schemes}
We also measured the order of convergence for the explicit Euler and Velocity Verlet integration schemes in a unit test (\texttt{integration\_schemes\_accuracy}). This is an artificial problem with a known analytical solution which is very similar to the collisions in DEM, and we use this test to investigate the orders of convergence of the integration schemes \cite{blais2019experimental}. This test calculates and compares the errors with respect to the analytical solution of the explicit Euler and Velocity Verlet schemes in a pendulum oscillation problem. Table~\ref{table:integration_order} reports the errors and orders of convergence.  The calculated orders of convergence for the explicit Euler and the Velocity Verlet schemes are 1 and 2, respectively.

\begin{table}[!htbp]
\centering
	\begin{tabular}{ |c|c|c|c| } 
		\hline
		 & \texttt{Explicit Euler} & \texttt{Velocity Verlet} \\
		\hline
		Error for $\Delta t=\SI{0.1}{\second}$ & 0.01275 & 0.00010\\ 
		Error for $\Delta t=\SI{0.05}{\second}$ & 0.00634 & $2.63\times10^{-5}$ \\ 
		Error for $\Delta t=\SI{0.025}{\second}$ & 0.00316 & $6.57\times10^{-6}$ \\
		 Order of convergence & $\approx1$ & $\approx2$ \\ 
		\hline
	\end{tabular}
\caption{Errors and orders of convergence of explicit Euler and Velocity Verlet integration schemes for a pendulum oscillation model (\texttt{integration\_schemes\_accuracy} unit test)}
\label{table:integration_order}
\end{table}

\subsubsection{Normal and tangential forces during a collision}
In this section, we compare the normal and tangential forces of a particle-wall collision at a specified condition (reported in Table~\ref{table:contact_properties}) with simulation results of another code (cemfDEM) \cite{norouzi2016coupled}. Figure~\ref{fig:normal_force} shows the results of this comparison in a plot of the normal force over the normal overlap. The results show high accuracy (AAE = 2.98\%, 3.54\% and 3.52\% for $e_n$ = 0.5, 0.7 and 0.9, respectively) of Lethe-DEM compared to cemfDEM \cite{norouzi2017new, norouzi2016coupled}. Small discrepancies between the normal forces in Figure~\ref{fig:normal_force} arise from different non-linear contact models used in the codes. For instance, the coefficient and power of the normal overlap in the damping force are different.

\begin{table}[!htbp]
\centering
	\begin{tabular}{ |c|c|c|c| } 
		\hline
		\textbf{Property} & \textbf{Value} \\
		\hline
		$d_p$ & $\SI{2}{\milli \meter}$ \\ 
		 $\rho_p$ & $\SI{7850}{\kilogram\per\cubic\meter}$ \\ 
		 $Y_p$,  $Y_w$ & $\SI{200}{\giga \pascal}$ \\ 
		 $\nu_p$, $\nu_w$ & $0.3$ \\ 
		 $\mu_p$, $\mu_w$ & $0.3$ \\ 
		 $e_p$, $e_w$ & $0.5-0.9$ \\ 
		 Impact velocity & $\SI[parse-numbers=false]{0.1-2}{\meter\per\second}$ \\ 
		\hline
	\end{tabular}
\caption{DEM parameters of the normal and tangential force calculations in Figures~\ref{fig:normal_force} and ~\ref{fig:tangential_force}}
\label{table:contact_properties}
\end{table}

\input{Figs/Figure_normal_force}

With the DEM parameters reported in Table~\ref{table:contact_properties}, we calculate the contact tangential force at three different contact angles: 5$^{\circ}$, 25$^{\circ}$ and 65$^{\circ}$ (Figure~\ref{fig:tangential_force}). At contact angles of 25$^{\circ}$ and 65$^{\circ}$, the tangential force is limited by Coulomb's limit (Equation~\eqref{eq:coulomb_overlap}). Other research \cite{kruggel2008study, norouzi2016coupled} report similar trends of tangential force at various contact angles.

\input{Figs/Figure_tangential_force}

\subsection{Validation}
We validate Lethe-DEM with experimental results of a flat-bottomed silo (Table~\ref{table:validation_info}-a), a wedge-shaped silo (Table~\ref{table:validation_info}-b), and a rotating drum (Table~\ref{table:validation_info}-c) filled with granular material. We chose these systems since both the normal and tangential forces govern their dynamics. In other words, the normal and tangential force calculations and integration are tested in these validations.

\renewcommand{\arraystretch}{1.5}
\begin{table}[!htbp]
\centering
	\begin{tabular}{ |c|c|c|c| }  
		\hline
         \textbf{Geometry} & \textbf{Properties} & \textbf{Value} & \textbf{Case} \\
		\hline 
	\multirow{2}{*}{\includegraphics[scale=.09]{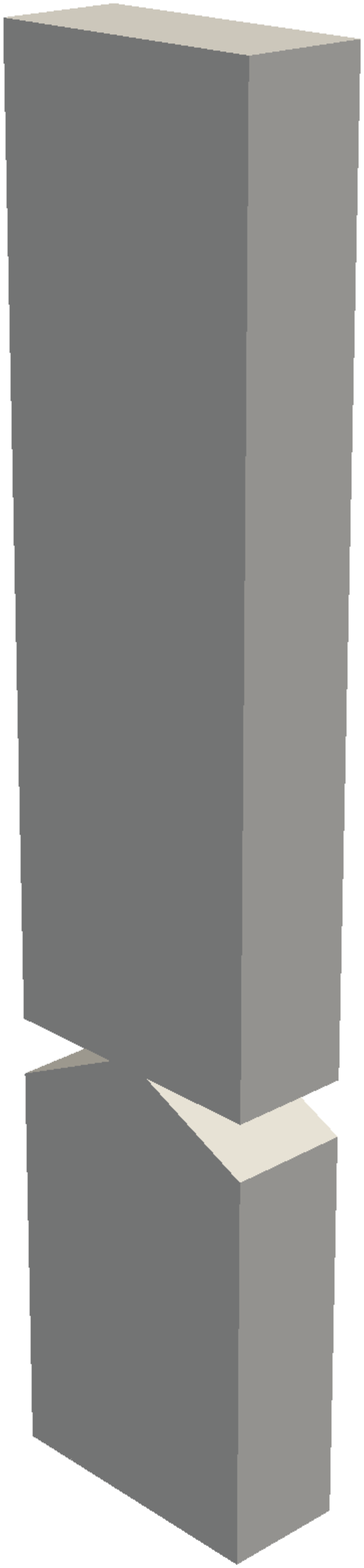}}&$d_p$ & $\SI{7.5}{\milli \meter}$ & \\
    &$\rho_p$ & $\SI{1290}{\kilogram\per\cubic\meter}$ &\\
    & $\nu_p$, $\nu_w$ & $0.2$ & (a)\\
    & $\mu_p$, $\mu_w$ & $0.29, 0.2$ & \\
    & $e_p$, $e_w$ & $0.56$ & \\
    & $\text{d}t$, $t_f$ & $10^{-5}\si{\second}$, $\SI{6}{\second}$ & \\
\hline
		\multirow{2}{*}{\includegraphics[scale=.1]{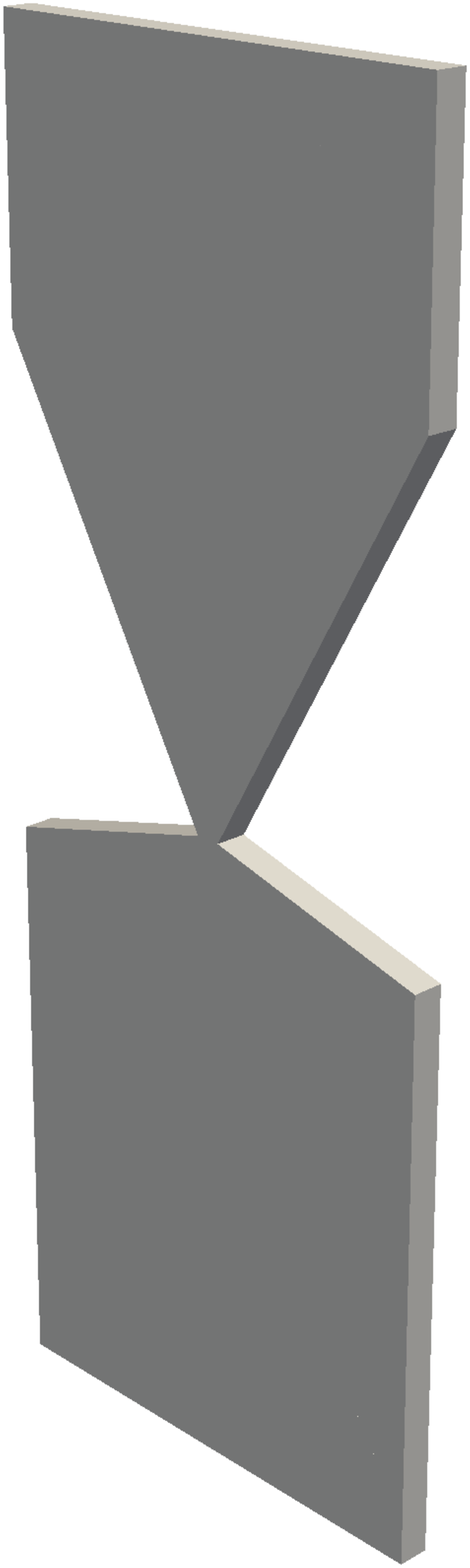}}&$d_p$ & $\SI{5.83}{\milli \meter}$ & \\
    &$\rho_p$ & $\SI{600}{\kilogram\per\cubic\meter}$ &\\
    & $\nu_p$, $\nu_w$ & $0.5$ &\\
    & $\mu_p$, $\mu_w$ & $0.5$ & (b) \\
    & $e_p$, $e_w$ & $0.7$ &\\
    & $f_{LB}$ & $\SI{0.5}{\hertz}$& \\
    & $\text{d}t$, $t_f$ & $10^{-5}\si{\second}$, $\SI{40}{\second}$& \\
\hline
		\multirow{2}{*}{\includegraphics[scale=.1]{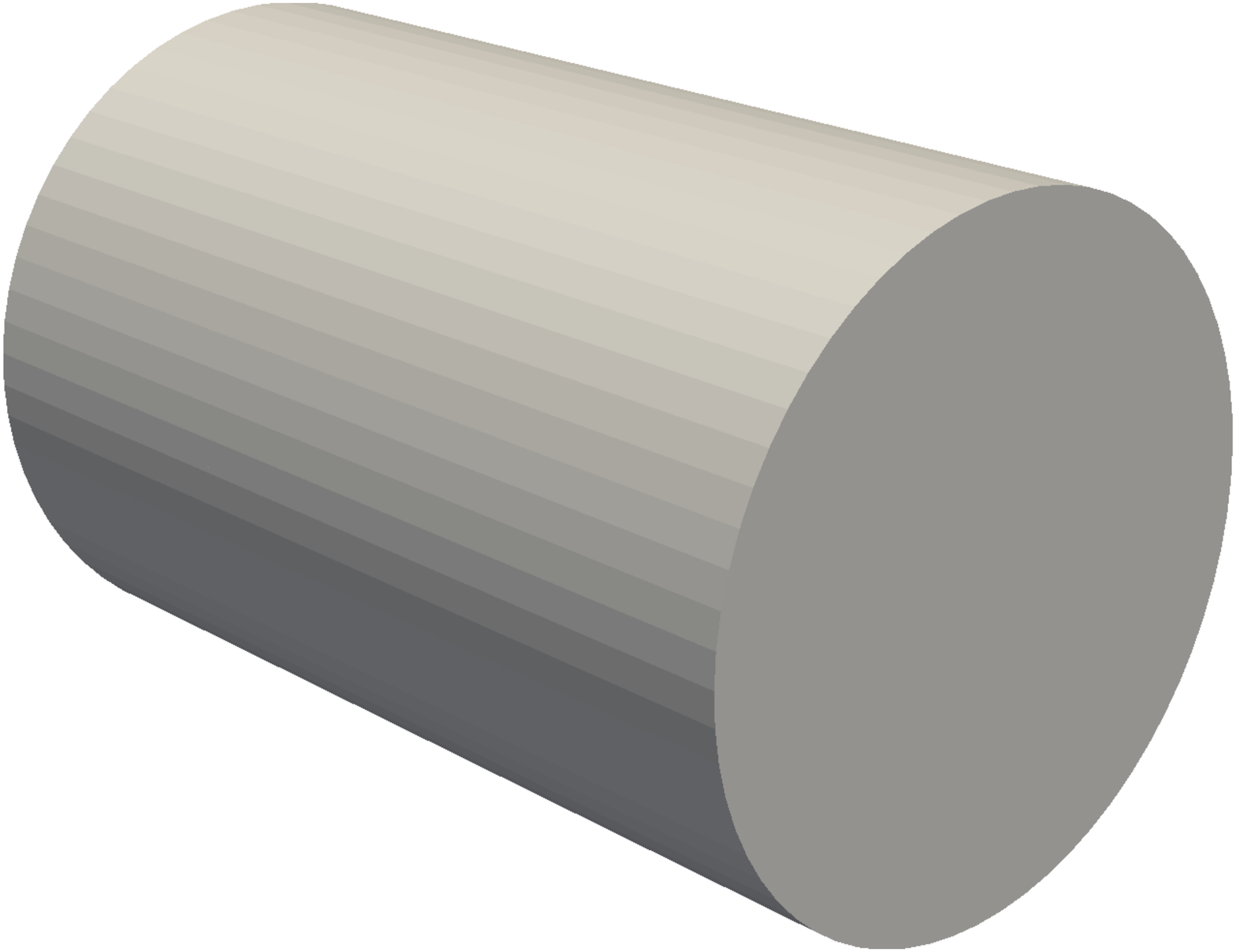}}&$d_p$ & $\SI{3}{\milli \meter}$ & \\
    &$\rho_p$ & $\SI{2500}{\kilogram\per\cubic\meter}$ & \\
    & $\nu_p$, $\nu_w$ & $0.2$ &  \\
    & $\mu_p$, $\mu_w$ & $0.85$  & (c)\\
    & $e_p$, $e_w$ & $0.97, 0.85$ & \\
    & $t_{LB}$ & $\SI{1.5}{\second}$ & \\
    & $\text{d}t$, $t_f$ & $10^{-5}\si{\second}$, $\SI{10}{\second}$ & \\
\hline
		\multirow{2}{*}{\includegraphics[scale=.1]{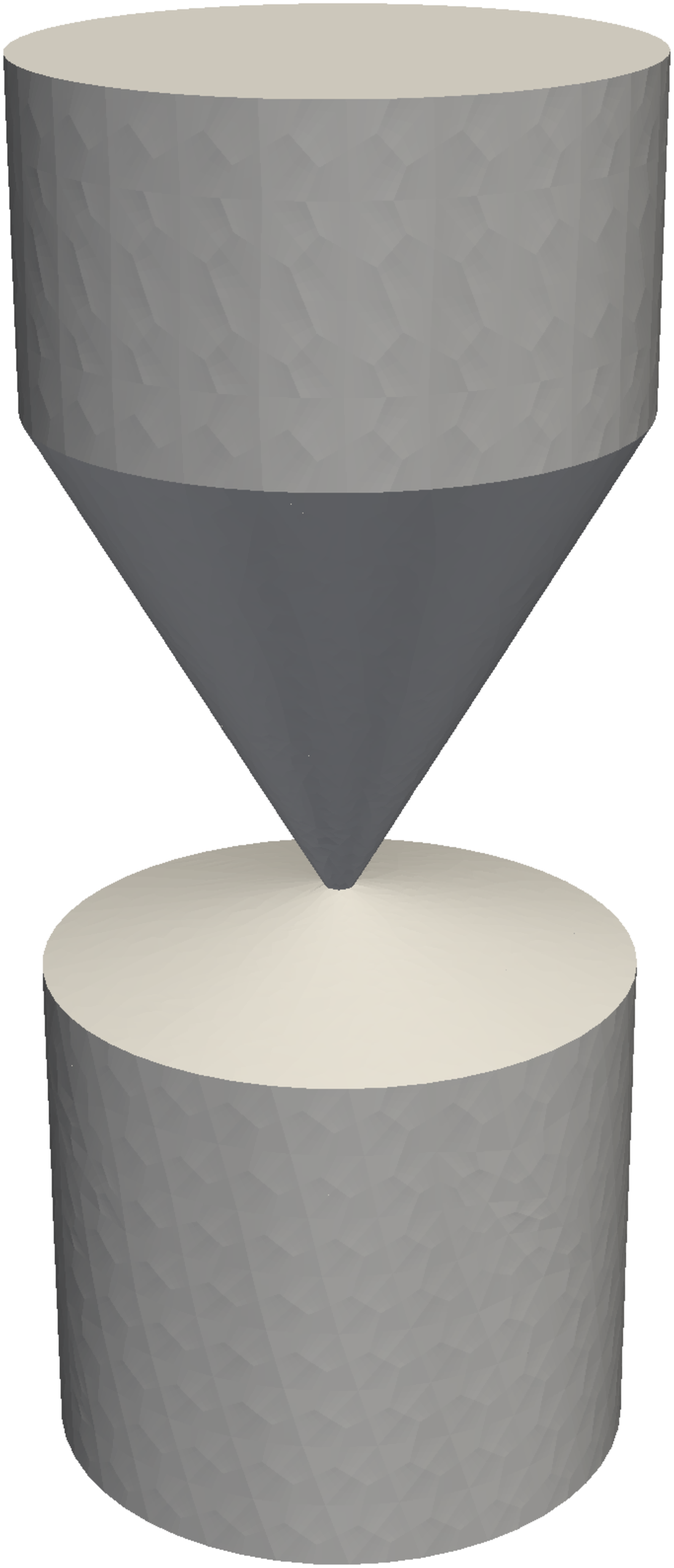}}&$d_p$ & $\SI{4}{\milli \meter}$ &\\
    &$\rho_p$ & $\SI{600}{\kilogram\per\cubic\meter}$ &\\
    & $\nu_p$, $\nu_w$ & $0.5$ & \\
    & $\mu_p$, $\mu_w$ & $0.5$ & (d) \\
    & $e_p$, $e_w$ & $0.7$ &\\
    & $f_{LB}$ & $\SI{2}{\hertz}$ &\\
    & $\text{d}t$, $t_f$ & $10^{-5}\si{\second}$, $\SI{6}{\second}$ &\\
		\hline
	\end{tabular}
\caption{Geometries of the validation and scalability simulations and the corresponding DEM parameters. (a) flat-bottomed silo, (b) wedge-shaped silo, (c) rotating drum, and (d) cylindrical silo}
\label{table:validation_info}
\end{table}

\subsubsection{Flat-bottomed silo}
Balevivcius et al. \cite{balevivcius2021experimental} measured the particle velocity using particle image velocimetry and DEM in a flat-bottomed silo with $n_p=18,000$ particles. Table~\ref{table:validation_info}-a shows the geometry of the silo and reports the physical properties of the simulation \cite{balevivcius2021experimental}. An animation (silo1.mp4) of this simulation is available in the supplementary materials. Figure~\ref{fig:silo1_validation} shows the lateral distributions of the vertical (Figure~\ref{fig:silo1_validation}-a) and lateral (Figure~\ref{fig:silo1_validation}-b) components of particle velocity in this flat-bottomed silo at three heights ($z=\SI{0.05}{\meter}$, $\SI{0.1}{\meter}$, and $\SI{0.15}{\meter}$). AAE values of the axial velocity distributions are 15.4\%, 25.6\% and 19.0\% at $z$= 0.05, 0.1 and 0.15, respectively. AAE values of lateral velocity distributions are also less than 17\%. Lethe-DEM predicts the trends and velocity magnitudes with acceptable accuracy. DEM results in the same work \cite{balevivcius2021experimental} have similar error values.  Errors are attributed to instantaneous experimental measurements. 

\input{Figs/Figure_silo1_validation}

\subsubsection{Wedge-shaped silo}\label{sec:second_silo}
Similar to the flat-bottomed silo, we compare the lateral distributions of axial and lateral components of particle velocity in a wedge-shaped silo with $n_p=\num{132300}$ particles. Golshan et al.~\cite{golshan2020experimental} measured time-averaged particle velocity in a wedge-shaped silo using particle image velocimetry (PIV) and DEM. Table~\ref{table:validation_info}-b reports the DEM physical properties of this simulation \cite{golshan2020experimental}. In the experiments, non-spherical barley grains were used, while in the DEM simulations we replaced them with spheres with the same volume as the non-spherical particles. Interested readers may find the simulation animation (silo2.mp4) in the supplementary materials. Figure~\ref{fig:silo2_validation} shows the lateral distributions of vertical and lateral components of particle velocity in this silo. AAE values of the comparison between DEM and experiments are less than 20\% for all the velocity distributions in Figure~\ref{fig:silo2_validation}. Errors are attributed to replacing non-spherical barley grain with spherical particles.

\begin{figure}[hbt!]
	\begin{center}
		\begin{tikzpicture}
			\begin{axis}[
				label style={font=\normalsize},
				legend style={nodes={scale=0.8, transform shape}},
				legend style={fill=none},
				xlabel={$x$},
				ylabel={$v_z$},
				x unit=\si{m}, y unit=m/s,
				width=7cm, height=5.25cm,
				xmin=-0.08, xmax=0.08,
				ymin=-0.22, ymax=-0.04,
				xtick={-0.08,-0.04,0,0.04,0.08},
				ytick={-0.2,-0.15,-0.1,-0.05,0},
				ticklabel style={
					/pgf/number format/fixed,
					/pgf/number format/precision=2
				}, scaled ticks=false,
				legend pos=south east
				]
				
				\addplot+[color=red, mark=o, only marks, mark size=3pt, error bars/.cd,y dir=both,y explicit] 
				coordinates {
					(	-0.0533022	,	-0.0693113	) +- (-0.0533022,0.005959245)
					(	-0.0355348,-0.0868075)+- (	-0.0355348,0.00898233372291667)
					(	-0.0177674	,	-0.126759	)+- (-0.0177674,0.018180494081)
					(	0	,	-0.180063	)+- (0,0.035423733969)
					(	0.0177674,	-0.175385	)+- (0.0177674,0.033682981558)
					(	0.0355348,	-0.0907928	) +- (0.0355348,0.00975654586517)
					(	0.0533022,-0.0672517)+- (0.0533022,0.0056436528195)			
				};
				
				\addplot[color=red, smooth, dashed]
				coordinates {
					(-0.0533022,-0.075977) 
					(-0.0355348,-0.09639) 
					(-0.0177674,-0.14347) 
					(0,-0.21161) 
					(0.0177674,-0.141985)
					(0.0355348,-0.09845) 
					(0.0533022,-0.071142) 			
				};
				
				\addplot+[color=blue, mark=square, only marks, , mark size=3pt, error bars/.cd,y dir=both,y explicit]
				coordinates {
					(-0.0710696,-0.0587446) +- (	-0.0710696	,	0.004430005	)
					(-0.0533022,-0.0762602) +- (	-0.0533022	,	0.007086621	)
					(-0.0355348,-0.0957724) +- (	-0.0355348	,	0.010768559	)
					(-0.0177674,-0.111574) +- (	-0.0177674	,	0.014308324	)
					(0,-0.119259) +- (	0	,	0.016210359	)
					(0.0177674,-0.121251) +- (	0.0177674	,	0.016722655	)
					(0.0355348,-0.0949846) +- (	0.0355348	,	0.010605151	)
					(0.0533022,-0.0722465) +- (	0.0533022	,	0.006423665	)
					(0.0710696,-0.0522749) +- (	0.0710696	,	0.003603914	)
									
				};
				
				\addplot[color=blue, smooth]
				coordinates {	
					(	-0.0710696	,	-0.047193	)
					(	-0.0533022	,	-0.071741	)
					(	-0.0355348	,	-0.09562	)
					(	-0.0177674	,	-0.1223423	)
					(	0	,	-0.13911	)
					(	0.0177674	,	-0.11918	)
					(	0.0355348	,	-0.0954	)
					(	0.0533022	,	-0.072142	)
					(	0.0710696	,	-0.04824124	)						
				};

				\legend{$z=0.05 \si{m}$,, $z=0.1 \si{m}$,, $z=0.1 \si{m}$}
				\node[draw,align=left,scale=0.9] (textbox) at (rel axis cs:0.08,0.12)  {a};
				
			\end{axis}
		\end{tikzpicture}
	
			\begin{tikzpicture}
		\begin{axis}[
			label style={font=\normalsize},
			legend style={nodes={scale=0.8, transform shape}},
			legend style={fill=none},
			xlabel={$x$},
			ylabel={$v_x$},
			x unit=\si{m}, y unit=m/s,
			width=7cm, height=5.25cm,
			xmin=-0.08, xmax=0.08,
			ymin=-0.07, ymax=0.07,
			xtick={-0.12,-0.08,-0.04,0,0.04,0.08,0.12},
			ytick={-0.06,-0.03,0,0.03,0.06},
			ticklabel style={
				/pgf/number format/fixed,
				/pgf/number format/precision=2
			}, scaled ticks=false,
			legend pos=north east
			]
			
			\addplot+[color=blue, mark=square, only marks , mark size=3pt, error bars/.cd,y dir=both,y explicit]
			coordinates {
			(-0.0533022	,0.0350611	) +- (	-0.0533022	,	0.004681954)
			(-0.0355348	,0.0463165	) +- (	-0.0355348	,	0.007922669	)
			(-0.0177674	,0.0380927	) +- (	-0.0177674	,	0.005471724	)
			(0	,0.00239966	) +- (	0	,	5.91889E-05	)
			(0.0177674	,-0.0339063	) +-(	0.0177674	,	0.004397229	)
			(0.0355348	,-0.0396584	) +- (	0.0355348	,	0.005903602	)
			(0.0533022	,-0.0417843	) +- (	0.0533022	,	0.006516164	)

			};
			
			\addplot[color=blue, smooth]
			coordinates {
				(	-0.0533022	,	0.047847	)
				(	-0.0355348	,	0.045723	)
				(	-0.0177674	,	0.02989	)
				(	0	,	0.002604	)
				(	0.0177674	,	-0.026104	)
				(	0.0355348	,	-0.047433	)
				(	0.0533022	,	-0.048359	)		
			};
			
			\addplot+[color=red, mark=o, only marks , mark size=3pt, error bars/.cd,y dir=both,y explicit]
			coordinates {
				(	-0.0710696	,	0.0349603	)+-	(	-0.0710696	,	0.003027117	)
				(	-0.0533022	,	0.0352923	)+-	(	-0.0533022	,	0.003079298	)
				(	-0.0355348	,	0.0345041	)+-	(	-0.0355348	,	0.002956134	)
				(	-0.0177674	,	0.0291098	)+-	(	-0.0177674	,	0.002179924	)
				(	0	,	0.0190572	)+-	(	0	,	0.001043974	)
				(	0.0177674	,	-0.000218317	)+-	(	0.0177674,3.73394E-06	)
				(	0.0355348	,	-0.0241808	)+-	(	0.0355348	,	0.001572436	)
				(	0.0533022	,	-0.0321102	)+-	(	0.0533022	,	0.0025973	)
				(	0.0710696	,	-0.0333662	)+-	(	0.0710696	,	0.00278271	)
				
			};
			
			\addplot[color=red, smooth, dashed]
			coordinates {
				(	-0.0710696	,	0.030758	)
				(	-0.0533022	,	0.0308591	)
				(	-0.0355348	,	0.029435	)
				(	-0.0177674	,	0.0234124	)
				(	0	,	0.0012533	)
				(	0.0177674	,	-0.020539	)
				(	0.0355348	,	-0.02874	)
				(	0.0533022	,	-0.031458	)
				(	0.0710696	,	-0.03084127	)
												
			};
			
			\legend{$z=0.05 \si{m}$,, $z=0.1 \si{m}$,, $z=0.1 \si{m}$}
			\node[draw,align=left,scale=0.9] (textbox) at (rel axis cs:0.08,0.12)  {b};
			
		\end{axis}
	\end{tikzpicture}
		
		\caption{Lateral distributions of (a) vertical, and (b) lateral components of particles velocity in a wedge-shaped silo, obtained from Lethe-DEM (lines) and experiments (markers) using PIV \cite{golshan2020experimental}.
			\label{fig:silo2_validation}}
	\end{center}
\end{figure}
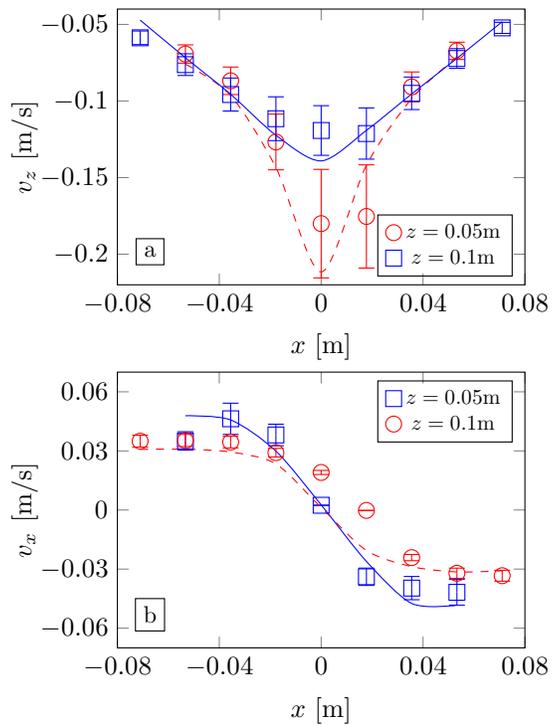

\begin{figure*} 
	\includegraphics[scale=.25]{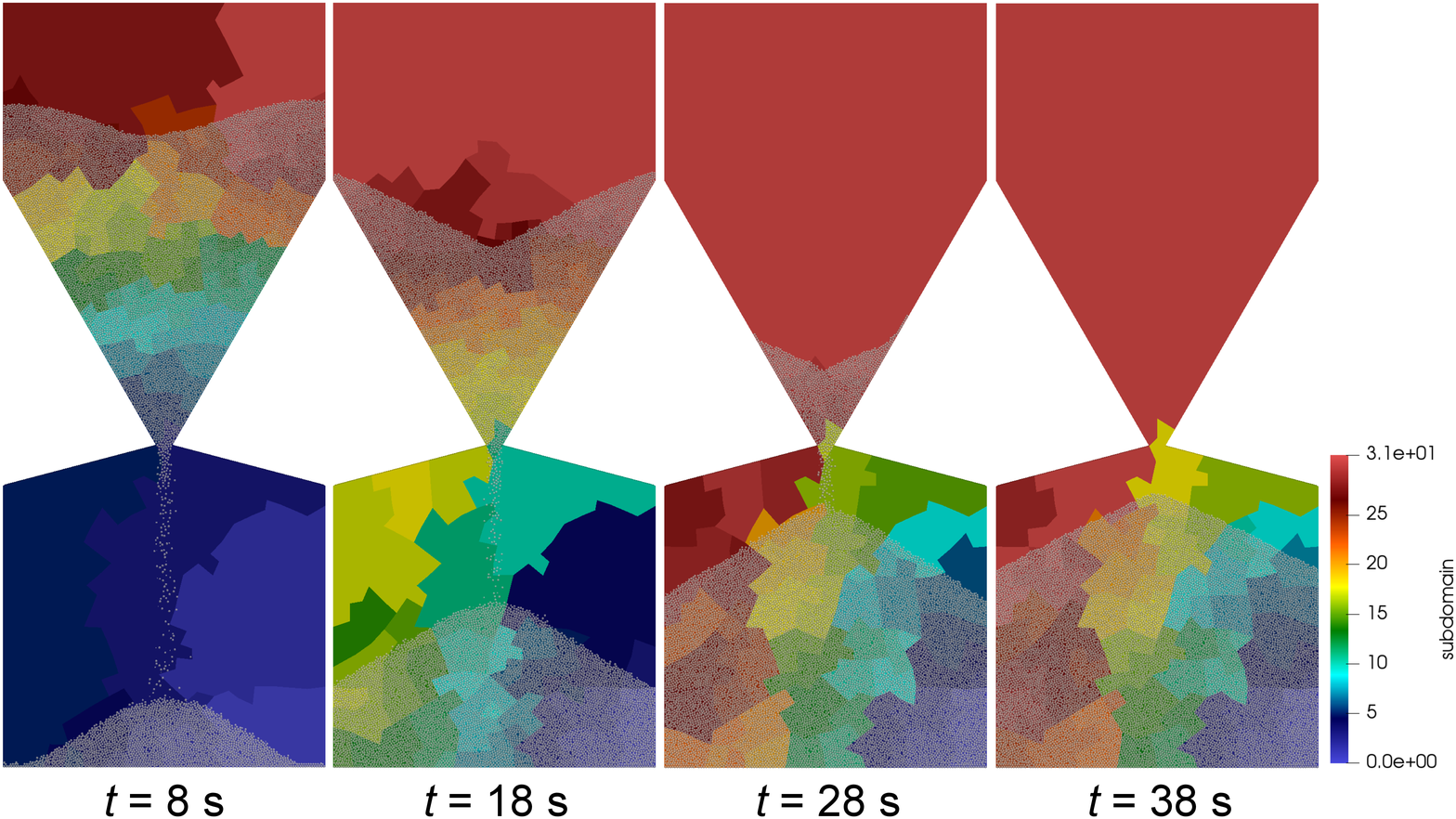}
	\centering
	\caption{Distribution of core domains as a function time in the simulation of wedge-shaped silo on 32 cores.
		\label{fig:LB_silo1}}
\end{figure*}

Load-balancing redistributes the computational load on the processes every \SI{2}{\second} in the wedge-shaped simulation. Figure~\ref{fig:LB_silo1} shows the distribution of core domains during the simulation (on 32 processes). At $t=\SI{8}{\second}$ (iteration $8\times10^6$), the majority of the cores (26 processes) are in the hopper section (upper part), and as the simulation proceeds the processes move toward the bottom of the geometry. At $t=\SI{38}{\second}$ (iteration $38\times10^6$), the particles are in the bottom of the container and only one process handles all the cells located in the hopper section.

\subsubsection{Rotating drum}
Here we compare the simulation results of a rotating drum containing $n_p=\num{226080}$ particles with the experiments of Alizadeh et al. \cite{alizadeh2013characterization}. Table~\ref{table:validation_info}-c reports the simulation parameters according to the physical properties of glass beads (particles) and plexiglass (wall). In the simulation of the wedge-shaped silo, we used a frequent load-balancing with the frequency of $f_{LB}=\SI{0.5}{\hertz}$ since the particles move from the hopper to the bottom container during the simulation. In the simulation of the rotating drum, after reaching the steady-state condition ($t\approx\SI{1.5}{\second}$), the particles occupy a specific part of the rotating cylinder, and the rest of the simulation domain is without any particles. From this point on, no load-balancing is performed anymore. As a result, we use a single-step load balance at $t=\SI{1.5}{\second}$, when all the particles have been inserted and have reached the particle bed.

\begin{figure*} 
	\includegraphics[scale=.3]{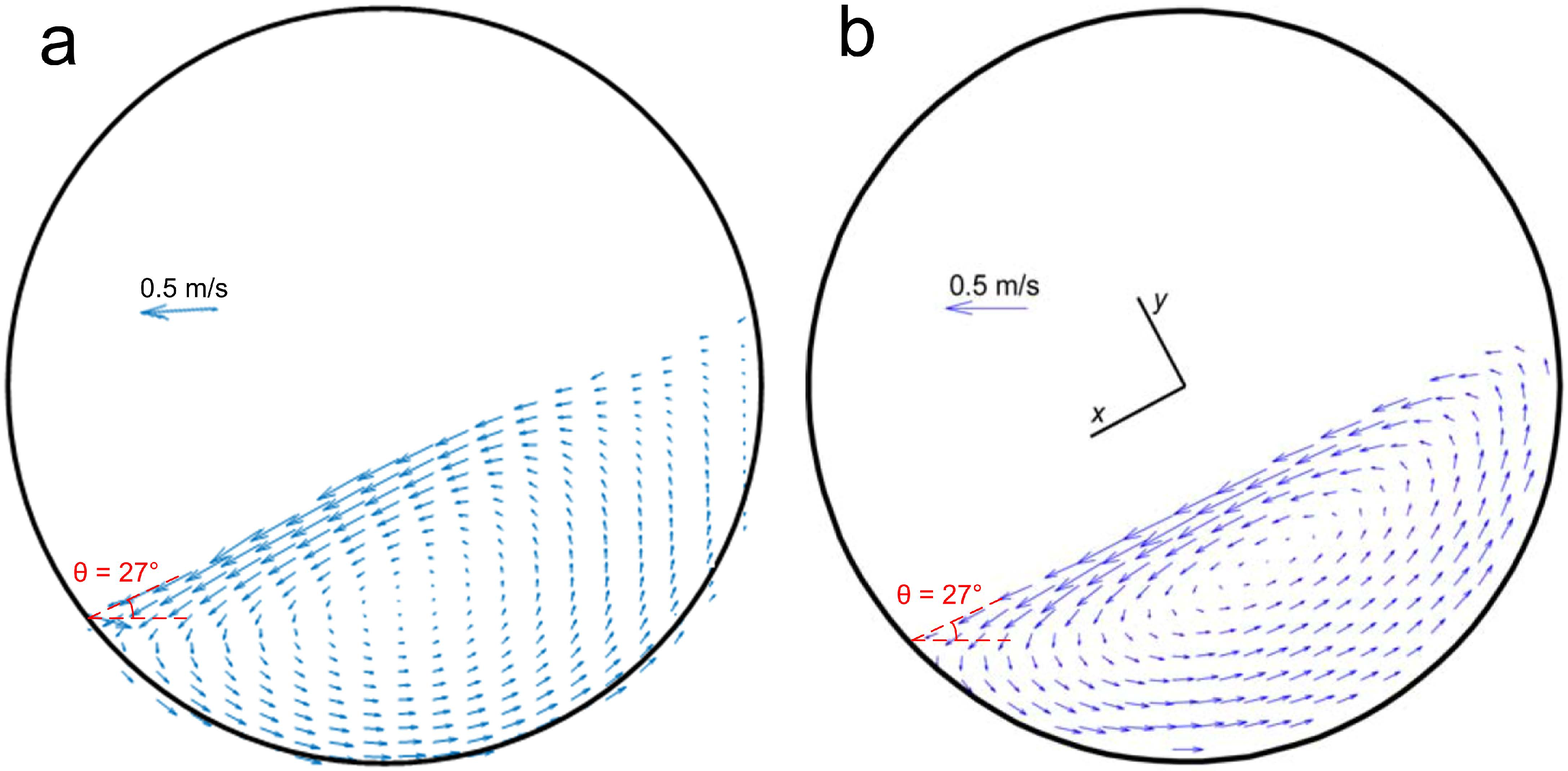}
	\centering
	\caption{Time-averaged particles velocity field in a rotating drum from (a) Lethe-DEM simulation and (b) experiments \cite{alizadeh2013characterization}. The free surface angle is equal to $27^{\circ}$ from the simulation and experiments.
		\label{fig:drum_velocity_field}}
\end{figure*}

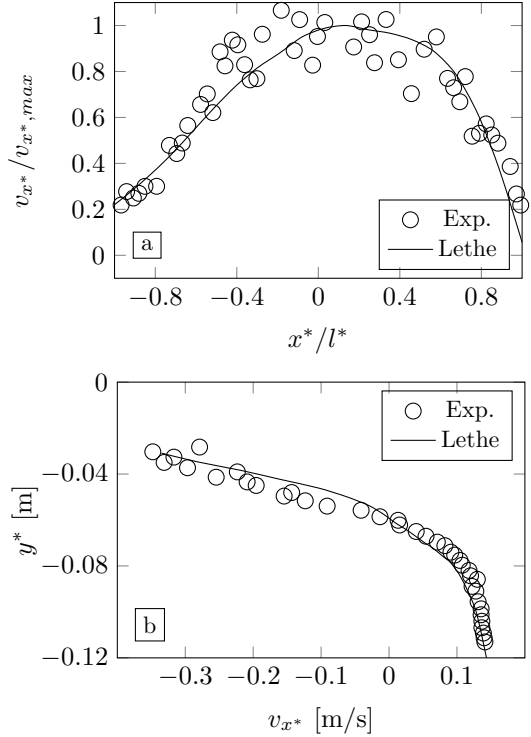
\begin{figure}[hbt!]
	\begin{center}
		\begin{tikzpicture}
		\begin{axis}[
		label style={font=\normalsize},
		legend style={nodes={scale=0.9, transform shape}},
		legend style={fill=none},
		xlabel={$x^*/l^*$},
		ylabel={$v_{x^*}/v_{x^*,max}$},
		width=7cm, height=5.25cm,
		xmin=-1, xmax=1,
		ymin=-0.1, ymax=1.1,
		xtick={-0.8,-0.4,0,0.4,0.8},
		ytick={0,0.2,0.4,0.6,0.8,1},
		ticklabel style={
			/pgf/number format/fixed,
			/pgf/number format/precision=2
		}, scaled ticks=false,
		legend pos=south east
		]
		
		\addplot[color=black, mark=o, only marks, mark size=3pt]
		coordinates {
			(	-0.967713878826689	,	0.218689248473147	)
			(	-0.940792848245035	,	0.277198046393378	)
			(	-0.907640156619865	,	0.249274087950528	)
			(	-0.880623548414845	,	0.269647414147399	)
			(	-0.85062491836078	,	0.300194022575434	)
			(	-0.793482243270538	,	0.300266661569193	)
			(	-0.694591262293671	,	0.442765251799248	)
			(	-0.667638372504229	,	0.488562225811692	)
			(	-0.730770578658791	,	0.478312481481835	)
			(	-0.64076194481348	,	0.564867577202825	)
			(	-0.517594247501441	,	0.620956349699408	)
			(	-0.577833637588767	,	0.656472994542517	)
			(	-0.544865729368773	,	0.702277614764831	)
			(	-0.457953810520547	,	0.824421994322688	)
			(	-0.482166808440141	,	0.885408164240587	)
			(	-0.42214406096578	,	0.936331921970428	)
			(	-0.395025503295835	,	0.916027411662381	)
			(	-0.361726259314835	,	0.829629063243713	)
			(	-0.33449300849685	,	0.763561986867634	)
			(	-0.301423150811931	,	0.76868877058503	)
			(	-0.274839827832841	,	0.9619429019278	)
			(	-0.181868287663159	,	1.06629837422462	)
			(	-0.118270937074878	,	0.890955489500797	)
			(	-0.027886364577658	,	0.827511063109904	)
			(	-0.004138511091783	,	0.952117522245691	)
			(	-0.091541061739959	,	1.02573523086774	)
			(	0.03179867529414	,	1.01318015426228	)
			(	0.173419225757532	,	0.906580519368805	)
			(	0.212242856369133	,	1.01595190533992	)
			(	0.332517737613936	,	1.02627428866353	)
			(	0.25148065668199	,	0.96006958050981	)
			(	0.275846578798971	,	0.838066655834535	)
			(	0.39310757898694	,	0.850927580834774	)
			(	0.519308272880486	,	0.896850717310063	)
			(	0.57932464851329	,	0.95031683982146	)
			(	0.456634839318085	,	0.703550708708077	)
			(	0.633912215138858	,	0.769877756219713	)
			(	0.721110866857185	,	0.777615720607491	)
			(	0.664089256756541	,	0.729238150764142	)
			(	0.694317273106687	,	0.668259627056113	)
			(	0.823882299342744	,	0.571814158869125	)
			(	0.754843396064113	,	0.518336567042923	)
			(	0.790901647439634	,	0.531094268209926	)
			(	0.851070947269825	,	0.523543635963948	)
			(	0.881235245204392	,	0.487988760071491	)
			(	0.941640303172221	,	0.386370630907891	)
			(	0.972014871878196	,	0.266917717224043	)
			(	0.993188501374724	,	0.218639548108996	)
		};
		
		\addplot[color=black, smooth]
		coordinates {
			(	1	,	0.056436976779739	)
			(	0.891963109354414	,	0.364110742760077	)
			(	0.783926218708827	,	0.612898822000455	)
			(	0.675889328063241	,	0.786237257919262	)
			(	0.567852437417655	,	0.887310266137146	)
			(	0.459815546772069	,	0.938294631748005	)
			(	0.351778656126482	,	0.966480545776397	)
			(	0.243741765480896	,	0.981641624	)
			(	0.13570487483531	,	1	)
			(	0.027667984189723	,	0.987517727970269	)
			(	-0.080368906455863	,	0.9500247047169	)
			(	-0.188405797101449	,	0.88111315835501	)
			(	-0.296442687747036	,	0.817429541293905	)
			(	-0.404479578392622	,	0.737098772065796	)
			(	-0.512516469038208	,	0.640275851997488	)
			(	-0.620553359683795	,	0.535343307579091	)
			(	-0.728590250329381	,	0.432157178057241	)
			(	-0.836627140974967	,	0.3410356069343	)
			(	-0.944664031620554	,	0.259130387834024	)
			(	-1.05270092226614	,	0.183712105480865	)			
		};

		\legend{Exp., Lethe}
		\node[draw,align=left,scale=0.9] (textbox) at (rel axis cs:0.08,0.12)  {a};
		
		\end{axis}
		\end{tikzpicture}
		
		\begin{tikzpicture}
		\begin{axis}[
		label style={font=\normalsize},
		legend style={nodes={scale=0.9, transform shape}},
		legend style={fill=none},
		xlabel={$v_{x^*}$},
		ylabel={$y^*$},
		x unit=\si{m/s}, y unit=m,
		width=7cm, height=5.25cm,
		xmin=-0.4, xmax=0.2,
		ymin=-0.12, ymax=0,
		xtick={-0.3,-0.2,-0.1,0,0.1},
		ytick={-0.12,-0.08,-0.04,0},
		ticklabel style={
			/pgf/number format/fixed,
			/pgf/number format/precision=2
		}, scaled ticks=false,
		legend pos=north east
		]
		
				\addplot[color=black, mark=o, only marks, mark size=3pt]
		coordinates {
			(	0.141111111111111	,	-0.11306209850107	)
			(	0.14	,	-0.111263383297644	)
			(	0.138888888888888	,	-0.109207708779443	)
			(	0.136666666666666	,	-0.106895074946466	)
			(	0.136666666666666	,	-0.10406852248394	)
			(	0.135555555555555	,	-0.101241970021413	)
			(	0.135555555555555	,	-0.098672376873662	)
			(	0.131111111111111	,	-0.09558886509636	)
			(	0.127777777777777	,	-0.090963597430407	)
			(	0.122222222222222	,	-0.088907922912206	)
			(	0.13	,	-0.085824411134904	)
			(	0.12	,	-0.084282655246253	)
			(	0.117777777777777	,	-0.081970021413276	)
			(	0.107777777777777	,	-0.0796573875803	)
			(	0.104444444444444	,	-0.077601713062099	)
			(	0.096666666666667	,	-0.075289079229122	)
			(	0.091111111111111	,	-0.074004282655246	)
			(	0.082222222222222	,	-0.071177730192719	)
			(	0.071111111111111	,	-0.069635974304069	)
			(	0.054444444444444	,	-0.067066381156317	)
			(	0.04	,	-0.065010706638116	)
			(	0.015555555555556	,	-0.062184154175589	)
			(	0.013333333333333	,	-0.060128479657388	)
			(	-0.013333333333333	,	-0.058586723768737	)
			(	-0.041111111111111	,	-0.05576017130621	)
			(	-0.091111111111111	,	-0.053961456102784	)
			(	-0.123333333333333	,	-0.051648822269807	)
			(	-0.143333333333333	,	-0.048051391862955	)
			(	-0.154444444444444	,	-0.049593147751606	)
			(	-0.195555555555555	,	-0.044967880085653	)
			(	-0.208888888888888	,	-0.043426124197002	)
			(	-0.223333333333333	,	-0.039057815845824	)
			(	-0.254444444444444	,	-0.041370449678801	)
			(	-0.296666666666666	,	-0.037259100642398	)
			(	-0.278888888888888	,	-0.028265524625268	)
			(	-0.316666666666666	,	-0.032633832976445	)
			(	-0.331111111111111	,	-0.034946466809422	)
			(	-0.347777777777777	,	-0.030321199143469	)
			
		};
		
		\addplot[color=black, smooth]
		coordinates {	
			(	0.143411305071639	,	-0.12	)
			(	0.127625362293538	,	-0.0993	)
			(	0.106805288901363	,	-0.0855	)
			(	0.091024068549227	,	-0.0787	)
			(	0.079613474445294	,	-0.0759	)
			(	0.064028640003259	,	-0.0721	)
			(	0.054295303898166	,	-0.0703	)
			(	0.0410975117818	,	-0.0675	)
			(	0.032720831618438	,	-0.0657	)
			(	0.023810604966997	,	-0.0639	)
			(	0.013726084920257	,	-0.0621	)
			(	0.004808940945661	,	-0.0603	)
			(	-0.008011396599231	,	-0.0575	)
			(	-0.029774264245321	,	-0.0537	)
			(	-0.062265764200237	,	-0.0499	)
			(	-0.103683704067356	,	-0.0461	)
			(	-0.160555305456855	,	-0.0423	)
			(	-0.216488013159372	,	-0.0385	)
			(	-0.281957069663122	,	-0.0347	)
			(	-0.334908076992431	,	-0.0309	)
		};
		
		\legend{Exp., Lethe}
		\node[draw,align=left,scale=0.9] (textbox) at (rel axis cs:0.08,0.12)  {b};
		
		\end{axis}
		\end{tikzpicture}
		
		\caption{Streamwise velocity profiles in (a) $x$ (parallel to the free surface) and (b) $y$ (perpendicular to the free surface) directions.
			\label{fig:drum_validation}}
	\end{center}
\end{figure}

Figure \ref{fig:drum_velocity_field} shows the time-averaged velocity field obtained from the simulation and experiments \cite{alizadeh2013characterization}. We also compare the free surface angle with a horizontal line on this figure. Figure \ref{fig:drum_validation} compares the streamwise velocity profiles in $x$ and $y$ directions (these directions are defined in Figure \ref{fig:drum_velocity_field}) between Lethe-DEM and experiments carried out using radioactive particle tracking \cite{alizadeh2013characterization}. These profiles show particles velocity in (a) parallel and (b) perpendicular planes to the free surface. AAE values of these profiles are less than $20\%$. Interested readers may find an animation of this simulation (drum.mp4) in the supplementary materials.

\subsection{Scalability}
In this section, we present the results of strong and weak scaling analyses. For the strong scaling, we use the simulations of the wedge-shaped silo (Table~\ref{table:validation_info}-b) and the rotating drum (Table~\ref{table:validation_info}-c), while for the weak scaling analysis, we use a dam break case. We used the Graham cluster for the strong scaling analysis on the wedge-shaped silo as well as the Cedar cluster for strong scaling analysis on the rotating drum and weak scaling analysis. Each node on both clusters consists of 32 cores in the form of 2 x Intel E5-2683 v4 Broadwell @ \SI{2.1}{\giga\hertz} processors with an available memory of 128 GiB per node.

\subsubsection{Strong scaling}
Strong scaling studies are not prevalent for DEM codes, since DEM codes generally fail to provide high speed-up values on large core numbers for constant problem size. For Lethe-DEM however, due to the unique design of the software, we report strong scaling analyses for the two cases investigated with and without load-balancing. To the authors' best knowledge, these results are amongst the few strong scaling analysis of DEM softwares.

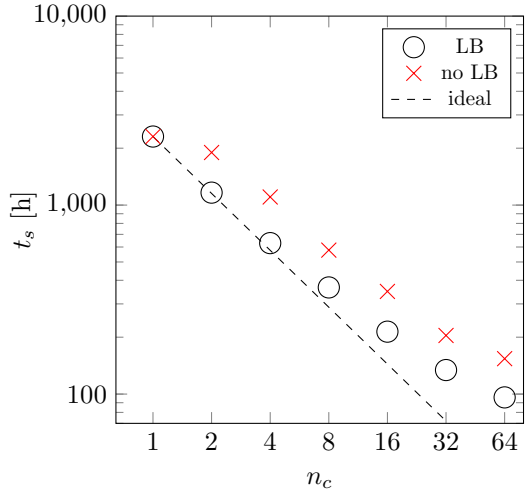
\begin{figure}[hbt!]
	\begin{center}
		\begin{tikzpicture}
		\begin{axis}[
		label style={font=\normalsize},
		legend style={nodes={scale=0.8, transform shape}},
		legend style={fill=none},
		xlabel={$n_c$},
		ylabel={$t_s$},
		width=7cm, height=7cm,
		xmin=0, xmax=80,
		ymin=70, ymax=10000,
		y unit=\si{\hour},
		xtick={1,2,4,8,16,32,64.01},
		ytick={10,100, 1000, 10000},
			xmode=log,	 ymode=log,
scaled ticks=false,
				     x tick label style={/pgf/number format/precision=0 sep=\,},
	log base 10 number format code/.code={%
	$\pgfmathparse{10^(#1)}\pgfmathprintnumber{\pgfmathresult}$%
	},
		legend pos=north east
		]
		
		\addplot+[color=black, mark=o, only marks, mark size=4pt, error bars/.cd,y dir=both,y explicit] 
		coordinates {
		(	1	,	2305.288	)
		(	2	,	1163.834	)
		(	4	,	629.14	)
		(	8	,	366.696	)
		(	16	,	213.925	)
		(	32	,	134	)
		(	64	,	96.042	)		
		};
		
		\addplot+[color=red, mark=x, only marks, mark size=4pt, error bars/.cd,y dir=both,y explicit] 
		coordinates {
		(	1	,	2305.288	)
		(	2	,	1897.981	)
		(	4	,	1103.800	)
		(	8	,	577.766	)
		(	16	,	349.286	)
		(	32	,	204.242	)
		(	64	,	153.953	)		
		};
		
		\addplot[color=black, smooth, dashed]
		coordinates {
				(	1	,	2305.288	)
		(	2	,	1152.644	)
		(	4	,	576.322	)
		(	8	,	288.161	)
		(	16	,	144.080	)
		(	32	,	72.04	)
		(	64	,	36.02	)
		(	128	,	18.01	)
		};

		\legend{LB, no LB, ideal}

		\end{axis}
		\end{tikzpicture}
		
		\caption{Simulation times of the wedge-shaped silo with and without load-balancing.
			\label{fig:silo2_cpu}}
	\end{center}
\end{figure}

As a first test, we consider a wedge-shaped silo on 1, 2, 4, 8, 16, 32, and 64 cores (i.e., using up to two nodes), and measure the simulation time, keeping the number of particles and the size of the domain fixed. The simulation of wedge-shaped silo calls load balancing every \SI{2}{\second} (i.e. every $2\times10^6$ steps) to distribute the computational load of the simulations on all the cores as particles move in the domain. Figure \ref{fig:silo2_cpu} shows the result of this comparison. 
For lower core counts ($n_c<16$) the strong scaling (with load-balancing) is quasi optimal. However, the speed-up deviates from the ideal line as the number of cores increases and the number of particles per core drops below 8,000 particles per core. This is because with increasing number of cores, the ratio of ghost particles to owned particles per process increases, which adds a non-negligible additional cost to the simulation. Furthermore, higher usage of shared resources like the memory bandwidth (up to 32 cores) or the network (64 cores) also leads to a deviation from the ideal behavior. Enabling load-balancing decreases the computational time by around $30\%$ depending on the number of processors. The simulation time per iteration decreases from \SI{2.07}{\second} to \SI{0.086}{\second} for this simulation on 1 to 64 processes with load-balancing.

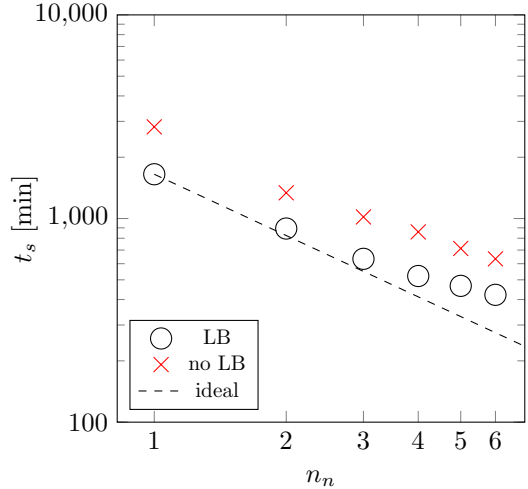
\begin{figure}[hbt!]
	\begin{center}
		\begin{tikzpicture}
		\begin{axis}[
		label style={font=\normalsize},
		legend style={nodes={scale=0.8, transform shape}},
		legend style={fill=none},
		xlabel={$n_n$},
		ylabel={$t_s$},
		y unit=\si{\minute},
		width=7cm, height=7cm,
		xmin=0, xmax=7,
		ymin=100, ymax=10000,
		xtick={1,2,3,4,5,6},
		ytick={100,1000,10000},
			xmode=log,	 ymode=log,
				     y tick label style={/pgf/number format/precision=0 sep=\,},
	log base 10 number format code/.code={%
	$\pgfmathparse{10^(#1)}\pgfmathprintnumber{\pgfmathresult}$%
	},
	legend pos=south west
		]
		
		\addplot+[color=black, mark=o, only marks, mark size=4pt, error bars/.cd,y dir=both,y explicit] 
		coordinates {
			(	1	,	1650	)
			(	2	,	893	)
			(	3	,	634	)
			(	4	,	522	)
			(	5	,	467	)
			(	6	,	422	)	
		};
		
				\addplot+[color=red, mark=x, only marks, mark size=4pt, error bars/.cd,y dir=both,y explicit] 
		coordinates {
			(	1	, 2819.28	)
			(	2	,	1337.26	)
			(	3	,	1018.23	)
			(	4	,	859.53	)
			(	5	,	712.88	)
			(	6	,	633.87)	
		};
		
		\addplot[color=black, smooth, dashed]
		coordinates {
			(	1	,	1650	)
			(	2	,	825	)
			(	3	,	550	)
			(	4	,	412.5	)
			(	7	,	235.714	)	
		};
		
		\legend{LB, no LB, ideal}
		
		\end{axis}
		\end{tikzpicture}
		
		\caption{Simulation time of rotating drum case as a function of number of nodes with and without load-balancing. Each node has 32 cores.
			\label{fig:drum_cpu}}
	\end{center}
\end{figure}

As a second test, we simulate the rotating drum on 1 to 6 compute nodes, each with 32 cores. Figure \ref{fig:drum_cpu} shows the simulation times of these simulations. Similar to the simulation of the wedge-shaped silo (Figure~\ref{fig:LB_silo1}), the deviation from the ideal speed-up increases with the number of cores. The best scalability is obtained at $\num{4000}<{n_p}/{n_c}<\num{20000}$ (1--3 nodes or 32--96 cores for the $~\num{226080}$ particles in this simulation). Using load-balancing decreases the computational time by around $35\%$. The simulation time per iteration decreases from \SI{0.099}{\second} to \SI{0.025}{\second} for this simulation on 1 to 6 nodes with load-balancing.

\subsubsection{Weak scaling}
\begin{figure} 
	\includegraphics[scale=.25]{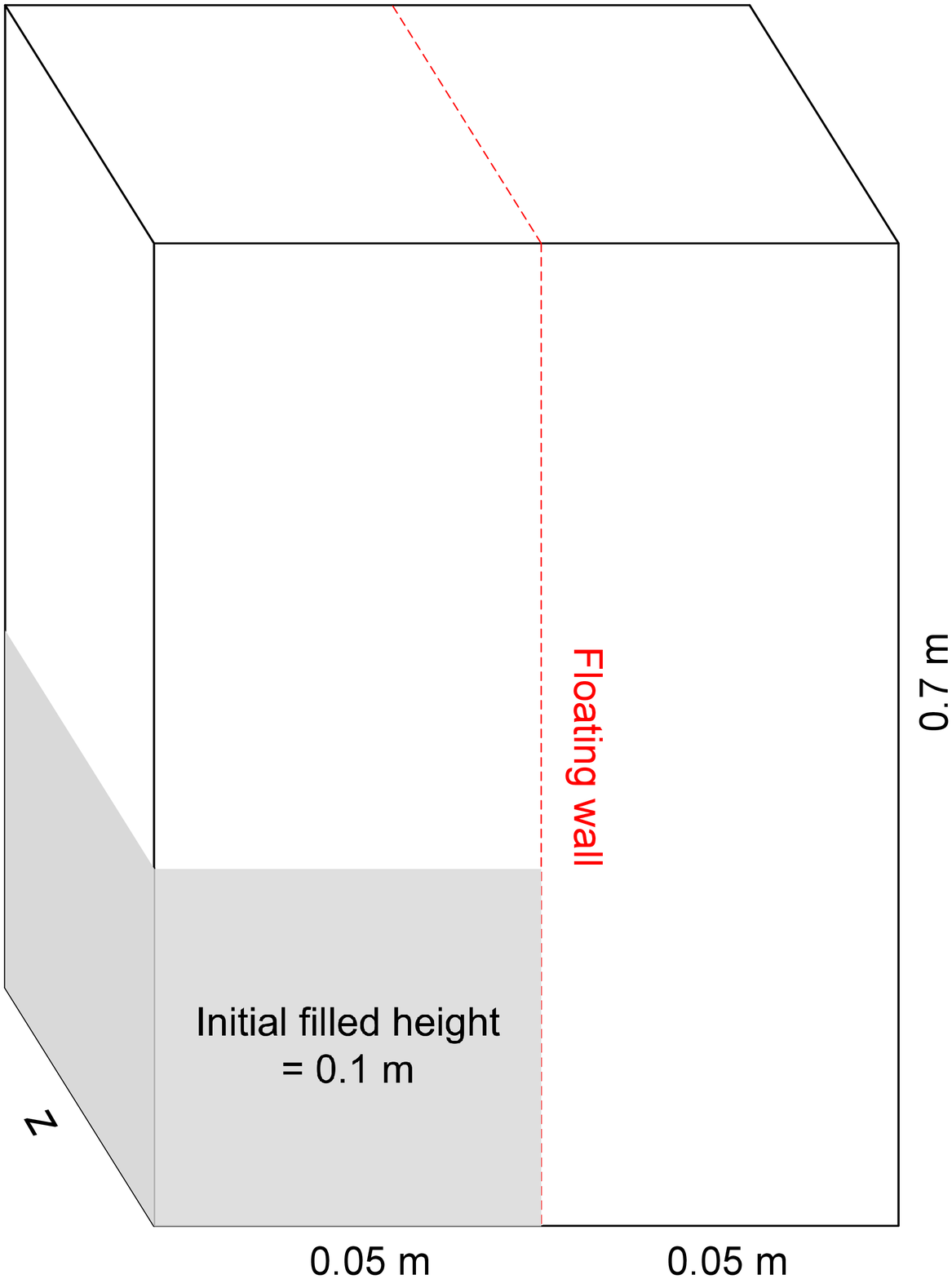}
	\centering
	\caption{Geometry of breaking of dam simulations used for weak-scaling anslysis. 
		\label{fig:weak_scale_geom}}
\end{figure}

We use dam break simulations for weak-scaling analysis of Lethe-DEM. In this case (illustrated in Figure~\ref{fig:weak_scale_geom}), particles are inserted and packed in the left half of a box (grey zone in Figure~\ref{fig:weak_scale_geom}). A floating wall dam (red line in Figure~\ref{fig:weak_scale_geom}) holds the particles in the left half. At $t=\SI{0.5}{\second}$, the dam is removed and particles fill the entire length of the box. We choose this scenario for weak-scaling analysis, since all the major functions of Lethe-DEM (including insertion, particle-particle and particle-wall contacts) are included and the particles can flow in the entire simulation domain. In weak-scaling analyses, the number of particles per process is kept constant. To this end, we increase the depth ($z$ in Figure~\ref{fig:weak_scale_geom}) of the box proportionally to the number of particles in the simulation. Consequently, particles owned by different processes are in contact. We use four values of $n_p/n_c$ : 5k, 10k, 15k and 20k. Table~\ref{table:weak_scaling} reports the values of $z$ and $n_p$ in weak-scaling simulations. Dynamic load-balancing redistributes the computational load between the processes during the weak-scaling simulations. In the dynamic load balancing approach, the code automatically detects the load-balancing steps from the distribution of particles and cells on the processes.

\begin{table}[!htbp]
\begin{tabular}{c|c|c|c|c|c|}
\cline{2-6}
    &\multicolumn{5}{c|}{\textbf{$n_p/n_c$}}                                                                       \\ \hline
\multicolumn{1}{|c|}{\multirow{9}{*}{\textbf{$n_c$}}} &              & \textbf{5k}                                                                           & \textbf{10k}                                                                          & \textbf{15k}                                                                          & \textbf{20k}                                                                          \\ \cline{2-6} 
\multicolumn{1}{|c|}{}                                & \textbf{32}  & \begin{tabular}[c]{@{}c@{}}$z=\SI{0.01}{\meter}$\\ $n_p=\SI{160}{k}$\end{tabular} & \begin{tabular}[c]{@{}c@{}}$z=\SI{0.02}{\meter}$\\ $n_p=\SI{320}{k}$\end{tabular} & \begin{tabular}[c]{@{}c@{}}$z=\SI{0.03}{\meter}$\\ $n_p=\SI{480}{k}$\end{tabular} & \begin{tabular}[c]{@{}c@{}}$z=\SI{0.04}{\meter}$\\ $n_p=\SI{640}{k}$\end{tabular} \\ \cline{2-6} 
\multicolumn{1}{|c|}{}                                & \textbf{64}  & \begin{tabular}[c]{@{}c@{}}$z=\SI{0.02}{\meter}$\\ $n_p=\SI{320}{k}$\end{tabular} & \begin{tabular}[c]{@{}c@{}}$z=\SI{0.04}{\meter}$\\ $n_p=\SI{640}{k}$\end{tabular} & \begin{tabular}[c]{@{}c@{}}$z=\SI{0.06}{\meter}$\\ $n_p=\SI{960}{k}$\end{tabular} & \begin{tabular}[c]{@{}c@{}}$z=\SI{0.08}{\meter}$\\ $n_p=\SI{1.28}{M}$\end{tabular}    \\ \cline{2-6} 
\multicolumn{1}{|c|}{}                                & \textbf{96}  & \begin{tabular}[c]{@{}c@{}}$z=\SI{0.03}{\meter}$\\ $n_p=\SI{480}{k}$\end{tabular} & \begin{tabular}[c]{@{}c@{}}$z=\SI{0.06}{\meter}$\\ $n_p=\SI{960}{k}$\end{tabular} & \begin{tabular}[c]{@{}c@{}}$z=\SI{0.09}{\meter}$\\ $n_p=\SI{1.44}{M}$\end{tabular}    & \begin{tabular}[c]{@{}c@{}}$z=\SI{0.12}{\meter}$\\ $n_p=\SI{1.92}{M}$\end{tabular}    \\ \cline{2-6} 
\multicolumn{1}{|c|}{}                                & \textbf{128} & \begin{tabular}[c]{@{}c@{}}$z=\SI{0.04}{\meter}$\\ $n_p=\SI{640}{k}$\end{tabular} & \begin{tabular}[c]{@{}c@{}}$z=\SI{0.08}{\meter}$\\ $n_p=\SI{1.28}{M}$\end{tabular}    & \begin{tabular}[c]{@{}c@{}}$z=\SI{0.12}{\meter}$\\ $n_p=\SI{1.92}{M}$\end{tabular}    & \begin{tabular}[c]{@{}c@{}}$z=\SI{0.16}{\meter}$\\ $n_p=\SI{2.56}{M}$\end{tabular}    \\ \cline{2-6} 
\multicolumn{1}{|c|}{}                                & \textbf{160} & \begin{tabular}[c]{@{}c@{}}$z=\SI{0.05}{\meter}$\\ $n_p=\SI{800}{k}$\end{tabular} & \begin{tabular}[c]{@{}c@{}}$z=\SI{0.1}{\meter}$\\ $n_p=\SI{1.6}{M}$\end{tabular}      & \begin{tabular}[c]{@{}c@{}}$z=\SI{0.15}{\meter}$\\ $n_p=\SI{2.4}{M}$\end{tabular}     & \begin{tabular}[c]{@{}c@{}}$z=\SI{0.2}{\meter}$\\ $n_p=\SI{3.2}{M}$\end{tabular}      \\ \cline{2-6} 
\multicolumn{1}{|c|}{}                                & \textbf{224} & \begin{tabular}[c]{@{}c@{}}$z=\SI{0.07}{\meter}$\\ $n_p=\SI{1.12}{M}$\end{tabular}    & \begin{tabular}[c]{@{}c@{}}$z=\SI{0.14}{\meter}$\\ $n_p=\SI{2.24}{M}$\end{tabular}    & \begin{tabular}[c]{@{}c@{}}$z=\SI{0.21}{\meter}$\\ $n_p=\SI{3.36}{M}$\end{tabular}    & \begin{tabular}[c]{@{}c@{}}$z=\SI{0.28}{\meter}$\\ $n_p=\SI{4.48}{M}$\end{tabular}    \\ \cline{2-6} 
\multicolumn{1}{|c|}{}                                & \textbf{256} & \begin{tabular}[c]{@{}c@{}}$z=\SI{0.08}{\meter}$\\ $n_p=\SI{1.28}{M}$\end{tabular}    & \begin{tabular}[c]{@{}c@{}}$z=\SI{0.16}{\meter}$\\ $n_p=\SI{2.56}{M}$\end{tabular}    & \begin{tabular}[c]{@{}c@{}}$z=\SI{0.24}{\meter}$\\ $n_p=\SI{3.84}{M}$\end{tabular}    & \begin{tabular}[c]{@{}c@{}}$z=\SI{0.32}{\meter}$\\ $n_p=\SI{5.12}{M}$\end{tabular}    \\ \cline{2-6} 
\multicolumn{1}{|c|}{}                                & \textbf{320} & \begin{tabular}[c]{@{}c@{}}$z=\SI{0.1}{\meter}$\\ $n_p=\SI{1.6}{M}$\end{tabular}      & \begin{tabular}[c]{@{}c@{}}$z=\SI{0.2}{\meter}$\\ $n_p=\SI{3.2}{M}$\end{tabular}      & \begin{tabular}[c]{@{}c@{}}$z=\SI{0.3}{\meter}$\\ $n_p=\SI{4.8}{M}$\end{tabular}      & \begin{tabular}[c]{@{}c@{}}$z=\SI{0.4}{\meter}$\\ $n_p=\SI{6.4}{M}$\end{tabular}      \\ \hline
\end{tabular}
\caption{Number of particles and depth of the box (illustrated in Figure~\ref{fig:weak_scale_geom}) used for weak-scaling analysis. $\si{\kilo}$ and $\si{M}$ stand for $\times10^3$ and $\times10^6$, respectively.}
\label{table:weak_scaling}
\end{table}

\begin{figure}[hbt!]
	\begin{center}
		\begin{tikzpicture}
		\begin{axis}[
		label style={font=\normalsize},
		legend style={nodes={scale=1, transform shape}},
		legend style={fill=none},
		xlabel={$n_n$},
		ylabel={$t_s$},
		y unit=min,
		width=9cm, height=9cm,
		xmin=0, xmax=11,
		ymin=0, ymax=1400,
		xtick={2,4, 6, 8, 10},
		ytick={0,300,600,900,1200},
		legend pos=north west,
		legend style={nodes={scale=0.85}}
		]
				\addplot[color=black,smooth, dashed, mark=square, mark options=solid, mark size=4pt, error bars/.cd, y dir=both,y explicit,error bar style=solid ]
		coordinates {
	(	1	,	736.9	)	+-	(	1	,	95.623	)
(	2	,	765.22	)	+-	(	2	,	32.491	)
(	3	,	802.28	)	+-	(	3	,	23.542	)
(	4	,	860.55	)	+-	(	4	,	62.523	)
(	5	,	890.92	)	+-	(	5	,	71.634	)
(	7	,	954.43	)	+-	(	7	,	49.001	)
(	8	,	974.5	)	+-	(	8	,	32.491	)
(	10	,	1010.7	)	+-	(	10	,	47.918	)
		};

		\addplot+[color=green, mark=o, mark size=4pt, smooth, error bars/.cd,y dir=both,y explicit] 
		coordinates {
(	1	,	141.1	)	+-	(	1	,	1.135	)
(	2	,	169.5	)	+-	(	2	,	8.25	)
(	3	,	191.48	)	+-	(	3	,	58.175	)
(	4	,	204.78	)	+-	(	4	,	1.78	)
(	5	,	228.1	)	+-	(	5	,	5.275	)
(	7	,	265.6	)	+-	(	7	,	6.905	)
(	8	,	298.5	)	+-	(	8	,	15.7	)
(	10	,	356.45	)	+-	(	10	,	4.59	)
		};
		
				\addplot+[color=black, mark=o, mark size=4pt, smooth, error bars/.cd,y dir=both,y explicit] 
		coordinates {
(	1	,	294.28	)	+-	(	1	,	34.065	)
(	2	,	317.07	)	+-	(	2	,	10.075	)
(	3	,	339.48	)	+-	(	3	,	24.37	)
(	4	,	370.9	)	+-	(	4	,	61.77	)
(	5	,	397.21	)	+-	(	5	,	51.533	)
(	7	,	441.1	)	+-	(	7	,	21.445	)
(	8	,	469.4	)	+-	(	8	,	15.335	)
(	10	,	530.72	)	+-	(	10	,	38.121	)
		};
		
		\addplot[color=blue, mark=o, mark size=4pt, smooth, error bars/.cd,y dir=both,y explicit]
		coordinates {
	    (	1	,	480.29	)	+-	(	1	,	53.97	)
(	2	,	512.65	)	+-	(	2	,	17.15	)
(	3	,	545.65	)	+-	(	3	,	21.9	)
(	4	,	581.73	)	+-	(	4	,	22.975	)
(	5	,	612.72	)	+-	(	5	,	3.145	)
(	7	,	670.1	)	+-	(	7	,	72.34	)
(	8	,	700.87	)	+-	(	8	,	1.5634	)
(   10  ,   728.22  )   +-  (   10  ,   32.573  )
		};
		
		\addplot[color=red, mark=o, mark size=4pt, smooth, error bars/.cd,y dir=both,y explicit]
		coordinates {
	(	1	,	689.42	)	+-	(	1	,	36.4	)
(	2	,	722.9	)	+-	(	2	,	52.114	)
(	3	,	761.87	)	+-	(	3	,	52.569	)
(	4	,	795.75	)	+-	(	4	,	81.991	)
(	5	,	822.52	)	+-	(	5	,	57.578	)
(   7   ,    850.08 )   +-  (   7   ,  80.554   )
(   8   ,    860.05  )   +-  (   8   ,  84.841  )
(   10  ,    863.83  )   +-  (   10  ,  46.254  )
		};

		\legend{\text{10k/core, no LB}, \text{5k/core}, \text{10k/core}, \text{15k/core},
		\text{20k/core},
	}
		
		\end{axis}
		\end{tikzpicture}
		
		\caption{Simulation times of weak-scaling analysis, at different values of $n_p/n_c$, illustrated in Table\ref{table:weak_scaling}. Each node has 32 cores. Error bars indicate the standard deviation across three runs of the same simulations.
			\label{fig:weak_scaling}}
	\end{center}
\end{figure}
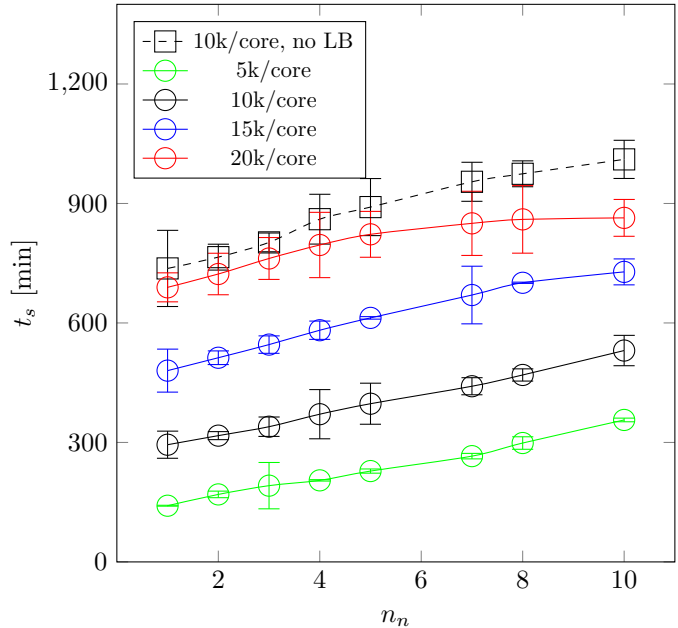

Figure~\ref{fig:weak_scaling} shows the results of the weak-scaling analysis by reporting the simulation times ($t_s$) against the number of nodes. The simulation times increase slightly with the number of processes at all the $n_p/n_c$ values. The ratio of simulation times on the maximum and minimum number of processes are 2.5 and 1.2 for $n_p/n_c=\text{5k}$ and $\text{20k}$, respectively. Note that for $n_p/n_c=\text{20k}$ on more than 6 nodes the simulation time does not further increase, which represents near ideal weak scalability in this range and suggests Lethe-DEM is well suited for large-scale parallel models. We also performed the weak-scaling analysis for $n_p/n_c= \text{10k}$ without load-balancing. We observe that the ratio of simulation times without and with load-balancing ($t_s^{\text{no LB}}/t_s^{\text{LB}}$) is $\approx2.4$ at $n_p/n_c=\text{10k}$ more or less independent of model size and therefore particle number. This means not only is the load-balancing very effective for this application, it is also scaling to larger problem sizes. The simulation time per iteration on 1--10 nodes changes from \SI{0.112}{\second} to \SI{0.285}{\second} for 5k/core and from \SI{0.578}{\second} to \SI{0.691}{\second} for \mbox{20k/core}. Interested readers may find a simulation animation (weak\_scaling.mp4) in the supplementary materials.

\subsection{Large scale feasibility}
\begin{figure} 
	\includegraphics[scale=.08]{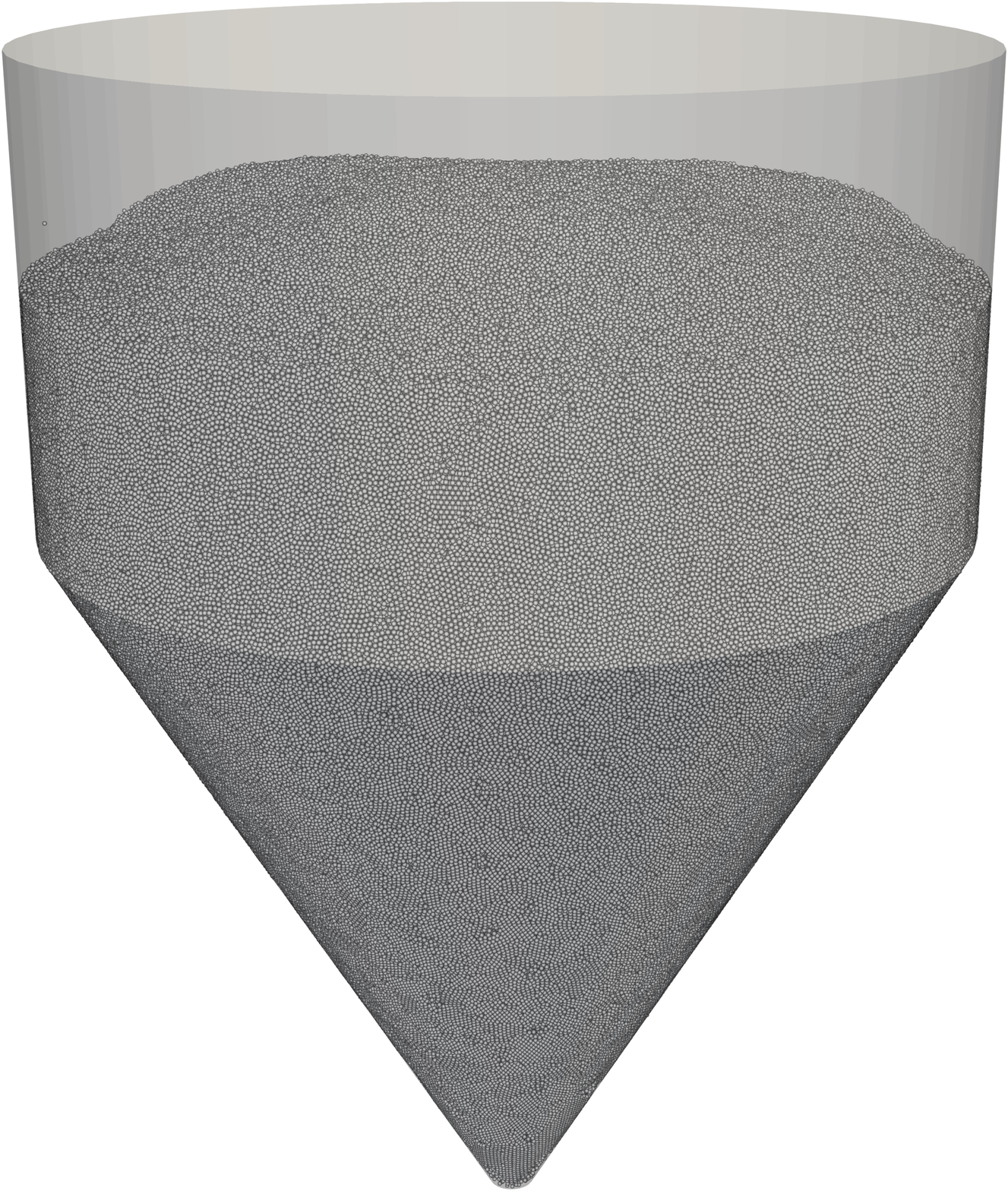}
	\centering
	\caption{Packing of 4.3M particles in a three-dimensional silo computed on 320 cores.
		\label{fig:large_silo}}
\end{figure}

We investigate the performance of Lethe-DEM in a simulation of a large system. To this end, we simulate the packing of $n_p>4.3\times10^6$ particles in a three-dimensional silo. The silo geometry is a three-dimensional (cylindrical) version of the silo used in Section~\ref{sec:second_silo}. Table~\ref{table:validation_info}-d reports the properties of the particles and simulation parameters. Figure~\ref{fig:large_silo} shows the packing configuration at $t=\SI{6}{\second}$. This simulation is performed on 320 cores and takes $\approx6$ days ($\approx\num{46800}$ core-hours) for $~\num{650000}$ time steps.

\section{Impact and conclusions}\label{section:conclusions}
In this work, we introduced an open-source parallel DEM software with load-balancing: Lethe-DEM. Lethe-DEM consists of two solvers for two-dimensional (\texttt{dem\_2d}) and three-dimensional (\texttt{dem\_3d}) simulations. We designed this software in the framework of Lethe, a high-order CFD solver for incompressible flows. Lethe-DEM uses a background grid that is compatible with other solvers in Lethe, which will allow coupling DEM and CFD simulations in the future. Several tests and comparisons with analytical solutions, models, and experiments verify and validate Lethe-DEM. For validation of Lethe-DEM, we compared its results with analogous experiments in a flat-bottomed silo, a wedge-shaped silo, and a rotating drum. Load-balancing allows Lethe-DEM to decrease the simulation times by roughly $\num{25}-\num{70}\%$ in parallel simulations compared to unbalanced parallel computations. To investigate the strong scalability of Lethe-DEM, we compared the simulation time of benchmark cases on 1--192 processes with and without load-balancing. Results show that Lethe-DEM scales optimally when there are $\approx \num{15000}$ particles per process, while scalability decreases for fewer particles per process due to increased communication costs. We also performed a weak-scaling analysis of Lethe-DEM using a dam break simulation on \num{32}--\num{320} processes and showed optimal weak scalability for more than $\approx \num{20000}$ particles per process. We finally demonstrated the computational performance and scalability of Lethe-DEM in the simulation of a cylindrical silo with $\num{4.3}\times10^6$ particles on \num{320} cores.

Lethe-DEM is under continuous development and we expect to add the following features in the near future:
\begin{itemize}
  \item Unresolved coupling with CFD \cite{berard2020experimental, golshan2020review};
  \item Heat transfer via conduction, convection and radiation mechanisms;
  \item Particle size change (particle size growth and shrinkage);
  \item Cohesive force models;
  \item Coarse graining;
  \item Electrostatic force.
\end{itemize}

The growing availability of large-scale massively parallel computing systems allows the modeling of physical processes at the pilot and industrial scales that can not be resolved by current simulations. However, in order to take advantage of this computing power we need DEM software that can leverage next generation HPC systems. Lethe-DEM was designed to use this opportunity and eventually enable coupled CFD-DEM models on the next generation of supercomputers.


\section{Acknowledgements}
The authors would like to acknowledge support received by the deal.II community. Without such rigorously developed open source project
such as deal.II, the present work could have never been achieved.
 The authors would like to acknowledge the support received from 
Calcul Qu\'ebec and Compute Canada. Computations shown in this work were made on the supercomputer Beluga, Cedar and Graham managed by Calcul Québec and Compute Canada. The operation of these supercomputers is funded by the Canada Foundation for Innovation (CFI), the ministère de l'Économie, de la science et de l'innovation du Québec (MESI) and the Fonds de recherche du Québec - Nature et technologies (FRQ-NT). This project was partially funded by the Natural Sciences and Engineering Research Council via NSERC Grant RGPIN-2020-04510.
RG was supported by the Computational Infrastructure for Geodynamics (CIG), through the National Science Foundation under Award No. EAR-1550901, administered by The University of California-Davis.

\section{Conflict of interest}
On behalf of all authors, the corresponding author states that there is no conflict of interest.

\appendix

\section{An example of the input files}

\singlespacing
\begin{lstlisting}[caption={Example of sections in a PRM file},label={lst::prm},columns=fullflexible,keepspaces=true,	breaklines=true,
frame=lines,	basicstyle=\small \ttfamily,
numberstyle=\scriptsize,
stepnumber=1,
numbersep=8pt,
showstringspaces=false,
breaklines=true,
frame=lines,
backgroundcolor=\color{background}]
# --------------------------------------------------
# Simulation and IO Control
#---------------------------------------------------
subsection simulation control
  set time step					    = 1e-5
  set time end					    = 0.4
  set log frequency				    = 1000000
  set output frequency				    = 1000000
end

#---------------------------------------------------
# Restart
#---------------------------------------------------
subsection restart
  set restart					    = true
  set checkpoint                                    = true
  set filename					    = sliding_restart
end

#---------------------------------------------------
# Timer
#---------------------------------------------------
subsection timer
  set type					    = none
end

#---------------------------------------------------
# Test
#---------------------------------------------------
subsection test
  set enable					    = true
end

# --------------------------------------------------
# Model parameters
#---------------------------------------------------
subsection model parameters
  set contact detection method 		   	    = constant
  set contact detection frequency                   = 20
  set neighborhood threshold			    = 1.3
  set particle particle contact force method        = pp_nonlinear
  set particle wall contact force method            = pw_nonlinear
  set rolling resistance torque method	  	    = constant_resistance
  set integration method			    = velocity_verlet
end

#---------------------------------------------------
# Physical Properties
#---------------------------------------------------
subsection physical properties
  set gx				    	    = 0.0
  set gy				    	    = 0.0
  set gz				    	    = -9.81
  set number of particle types			    = 1
  subsection particle type 0
    set size distribution type			    = uniform
    set diameter				    = 0.005
    set number					    = 20
    set density					    = 2000
    set young modulus particle			    = 1000000
    set poisson ratio particle			    = 0.3
    set restitution coefficient particle	    = 0.3
    set friction coefficient particle		    = 0.1
    set rolling friction particle		    = 0.05
  end
  set young modulus wall			    = 1000000
  set poisson ratio wall			    = 0.3
  set restitution coefficient wall          	    = 0.3
  set friction coefficient wall        		    = 0.1
  set rolling friction wall         	      	    = 0.05
end

#---------------------------------------------------
# Insertion Info
#---------------------------------------------------
subsection insertion info
  set insertion method				    = non_uniform
  set inserted number of particles at each time step= 20
  set insertion frequency            		    = 2000000
  set insertion box minimum x          		    = -0.05
  set insertion box minimum y          		    = -0.05
  set insertion box minimum z          		    = -0.06
  set insertion box maximum x          		    = 0.05
  set insertion box maximum y          		    = 0.05
  set insertion box maximum z          		    = 0
  set insertion distance threshold       	    = 2
  set insertion random number range           	    = 0.75
  set insertion random number seed           	    = 19
end

#---------------------------------------------------
# Mesh
#---------------------------------------------------
subsection mesh
  set type            		    		    = dealii
  set grid type            		            = hyper_cube
  set grid arguments            	    	    = -0.07:0.07:true
  set initial refinement            		    = 3
end

#---------------------------------------------------
# Boundary Motion
#---------------------------------------------------
subsection boundary motion
  set number of boundary motion        		    = 1
  subsection moving boundary 0
    set boundary id	 			    = 4
    set type              			    = translational
    set speed x					    = 0.15
    set speed y					    = 0
    set speed z					    = 0
  end
end
\end{lstlisting}

\bibliography{main}

\begin{thebibliography}{10}
\expandafter\ifx\csname url\endcsname\relax
  \def\url#1{\texttt{#1}}\fi
\expandafter\ifx\csname urlprefix\endcsname\relax\def\urlprefix{URL }\fi
\expandafter\ifx\csname href\endcsname\relax
  \def\href#1#2{#2} \def\path#1{#1}\fi

\bibitem{richard2005slow}
P.~Richard, M.~Nicodemi, R.~Delannay, P.~Ribiere, D.~Bideau, Slow relaxation
  and compaction of granular systems, Nature materials 4~(2) (2005) 121--128.

\bibitem{larsson2019particle}
S.~Larsson, Particle methods for modelling granular material flow, Ph.D.
  thesis, Lule{\aa} University of Technology (2019).

\bibitem{nedderman2005statics}
R.~M. Nedderman, Statics and kinematics of granular materials, Cambridge
  University Press, 2005.

\bibitem{golshan2019modeling}
S.~Golshan, R.~Zarghami, K.~Saleh, Modeling methods for gravity flow of
  granular solids in silos, Reviews in Chemical Engineering 1~(ahead-of-print)
  (2019).

\bibitem{blais2019experimental}
B.~Blais, D.~Vidal, F.~Bertrand, G.~S. Patience, J.~Chaouki, Experimental
  methods in chemical engineering: Discrete element method—dem, The Canadian
  Journal of Chemical Engineering 97~(7) (2019) 1964--1973.

\bibitem{golshan2020review}
S.~Golshan, R.~Sotudeh-Gharebagh, R.~Zarghami, N.~Mostoufi, B.~Blais,
  J.~Kuipers, Review and implementation of {CFD}-{DEM} applied to chemical
  process systems, Chemical Engineering Science (2020) 115646.

\bibitem{norouzi2016coupled}
H.~R. Norouzi, R.~Zarghami, R.~Sotudeh-Gharebagh, N.~Mostoufi, Coupled CFD-DEM
  modeling: formulation, implementation and application to multiphase flows,
  John Wiley \& Sons, 2016.

\bibitem{dan2018numerical}
H.-C. Dan, Z.~Zhang, J.-Q. Chen, H.~Wang, Numerical simulation of an indirect
  tensile test for asphalt mixtures using discrete element method software,
  Journal of Materials in Civil Engineering 30~(5) (2018) 04018067.

\bibitem{coetzee2017calibration}
C.~Coetzee, Calibration of the discrete element method, Powder Technology 310
  (2017) 104--142.

\bibitem{wolff2013three}
M.~F. Wolff, V.~Salikov, S.~Antonyuk, S.~Heinrich, G.~A. Schneider,
  Three-dimensional discrete element modeling of micromechanical bending tests
  of ceramic--polymer composite materials, Powder technology 248 (2013) 77--83.

\bibitem{tavarez2007discrete}
F.~A. Tavarez, M.~E. Plesha, Discrete element method for modelling solid and
  particulate materials, International journal for numerical methods in
  engineering 70~(4) (2007) 379--404.

\bibitem{cook2002discrete}
B.~K. Cook, R.~P. Jensen, Discrete element methods, American Society of Civil
  Engineers, 2002.

\bibitem{ismail2015discrete}
Y.~Ismail, Y.~Sheng, D.~Yang, J.~Ye, Discrete element modelling of
  unidirectional fibre-reinforced polymers under transverse tension, Composites
  Part B: Engineering 73 (2015) 118--125.

\bibitem{jerier2011study}
J.-F. Jerier, B.~Hathong, V.~Richefeu, B.~Chareyre, D.~Imbault, F.-V. Donze,
  P.~Doremus, Study of cold powder compaction by using the discrete element
  method, Powder Technology 208~(2) (2011) 537--541.

\bibitem{ketterhagen2009process}
W.~R. Ketterhagen, M.~T. am~Ende, B.~C. Hancock, Process modeling in the
  pharmaceutical industry using the discrete element method, Journal of
  pharmaceutical sciences 98~(2) (2009) 442--470.

\bibitem{tijskens2003discrete}
E.~Tijskens, H.~Ramon, J.~De~Baerdemaeker, Discrete element modelling for
  process simulation in agriculture, Journal of sound and vibration 266~(3)
  (2003) 493--514.

\bibitem{boac2014applications}
J.~M. Boac, R.~K. Ambrose, M.~E. Casada, R.~G. Maghirang, D.~E. Maier,
  Applications of discrete element method in modeling of grain postharvest
  operations, Food Engineering Reviews 6~(4) (2014) 128--149.

\bibitem{blais2016development}
B.~Blais, M.~Lassaigne, C.~Goniva, L.~Fradette, F.~Bertrand, Development of an
  unresolved cfd--dem model for the flow of viscous suspensions and its
  application to solid--liquid mixing, Journal of Computational Physics 318
  (2016) 201--221.

\bibitem{berard2020experimental}
A.~B{\'e}rard, G.~S. Patience, B.~Blais, Experimental methods in chemical
  engineering: Unresolved cfd-dem, The Canadian Journal of Chemical Engineering
  98~(2) (2020) 424--440.

\bibitem{kloss2012models}
C.~Kloss, C.~Goniva, A.~Hager, S.~Amberger, S.~Pirker, Models, algorithms and
  validation for opensource dem and cfd--dem, Progress in Computational Fluid
  Dynamics, an International Journal 12~(2-3) (2012) 140--152.

\bibitem{weinhart2020fast}
T.~Weinhart, L.~Orefice, M.~Post, M.~P. van Schrojenstein~Lantman, I.~F.
  Denissen, D.~R. Tunuguntla, J.~Tsang, H.~Cheng, M.~Y. Shaheen, H.~Shi,
  et~al., Fast, flexible particle simulations—an introduction to mercurydpm,
  Computer physics communications 249 (2020) 107129.

\bibitem{forgber2020extended}
T.~Forgber, P.~Toson, S.~Madlmeir, H.~Kureck, J.~G. Khinast, D.~Jajcevic,
  Extended validation and verification of xps/avl-fire™, a computational
  cfd-dem software platform, Powder Technology 361 (2020) 880--893.

\bibitem{blais2020lethe}
B.~Blais, L.~Barbeau, V.~Bibeau, S.~Gauvin, T.~El~Geitani, S.~Golshan,
  R.~Kamble, G.~Mikahori, J.~Chaouki, Lethe: An open-source parallel high-order
  adaptative cfd solver for incompressible flows, SoftwareX 12 (2020) 100579.

\bibitem{arndt2020deal}
D.~Arndt, W.~Bangerth, B.~Blais, T.~C. Clevenger, M.~Fehling, A.~V. Grayver,
  T.~Heister, L.~Heltai, M.~Kronbichler, M.~Maier, et~al., The deal. ii
  library, version 9.2, Journal of Numerical Mathematics 28~(3) (2020)
  131--146.

\bibitem{kronbichler2012}
M.~Kronbichler, K.~Kormann, A generic interface for parallel cell-based finite
  element operator application, Computers and Fluids 63 (2012) 135--147.

\bibitem{burstedde2011p4est}
C.~Burstedde, L.~C. Wilcox, O.~Ghattas, p4est: Scalable algorithms for parallel
  adaptive mesh refinement on forests of octrees, SIAM Journal on Scientific
  Computing 33~(3) (2011) 1103--1133.

\bibitem{burstedde2020}
C.~Burstedde, \href{https://doi.org/10.1145/3401990}{Parallel tree algorithms
  for amr and non-standard data access}, ACM Trans. Math. Softw. 46~(4) (Nov.
  2020).
\newblock \href {https://doi.org/10.1145/3401990} {\path{doi:10.1145/3401990}}.
\newline\urlprefix\url{https://doi.org/10.1145/3401990}

\bibitem{cundall1979discrete}
P.~A. Cundall, O.~D. Strack, A discrete numerical model for granular
  assemblies, geotechnique 29~(1) (1979) 47--65.

\bibitem{tsuji1992lagrangian}
Y.~Tsuji, T.~Tanaka, T.~Ishida, Lagrangian numerical simulation of plug flow of
  cohesionless particles in a horizontal pipe, Powder technology 71~(3) (1992)
  239--250.

\bibitem{zhou1999rolling}
Y.~Zhou, B.~Wright, R.~Yang, B.~H. Xu, A.-B. Yu, Rolling friction in the
  dynamic simulation of sandpile formation, Physica A: Statistical Mechanics
  and its Applications 269~(2-4) (1999) 536--553.

\bibitem{brilliantov1998rolling}
N.~V. Brilliantov, T.~P{\"o}schel, Rolling friction of a viscous sphere on a
  hard plane, EPL (Europhysics Letters) 42~(5) (1998) 511.

\bibitem{delacroix2020simulation}
B.~Delacroix, A.~Bouarab, L.~Fradette, F.~Bertrand, B.~Blais, Simulation of
  granular flow in a rotating frame of reference using the discrete element
  method, Powder Technology 369 (2020) 146--161.

\bibitem{geuzaine2009gmsh}
C.~Geuzaine, J.-F. Remacle, Gmsh: A 3-d finite element mesh generator with
  built-in pre-and post-processing facilities, International journal for
  numerical methods in engineering 79~(11) (2009) 1309--1331.

\bibitem{martin2010mastering}
K.~Martin, B.~Hoffman, Mastering CMake: a cross-platform build system, Kitware,
  2010.

\bibitem{gropp1996high}
W.~Gropp, E.~Lusk, N.~Doss, A.~Skjellum, A high-performance, portable
  implementation of the mpi message passing interface standard, Parallel
  computing 22~(6) (1996) 789--828.

\bibitem{bangerth2012}
W.~Bangerth, C.~Burstedde, T.~Heister, M.~Kronbichler,
  \href{https://doi.org/10.1145/2049673.2049678}{Algorithms and data structures
  for massively parallel generic adaptive finite element codes}, ACM Trans.
  Math. Softw. 38~(2) (Jan. 2012).
\newblock \href {https://doi.org/10.1145/2049673.2049678}
  {\path{doi:10.1145/2049673.2049678}}.
\newline\urlprefix\url{https://doi.org/10.1145/2049673.2049678}

\bibitem{gassmoller2018flexible}
R.~Gassm{\"o}ller, H.~Lokavarapu, E.~Heien, E.~G. Puckett, W.~Bangerth,
  Flexible and scalable particle-in-cell methods with adaptive mesh refinement
  for geodynamic computations, Geochemistry, Geophysics, Geosystems 19~(9)
  (2018) 3596--3604.

\bibitem{garg2012open}
R.~Garg, J.~Galvin, T.~Li, S.~Pannala, Open-source mfix-dem software for
  gas--solids flows: Part i—verification studies, Powder Technology 220
  (2012) 122--137.

\bibitem{norouzi2017new}
H.~R. Norouzi, R.~Zarghami, N.~Mostoufi, New hybrid {CPU}-{GPU} solver for
  {CFD}-{DEM} simulation of fluidized beds, Powder Technology 316 (2017)
  233--244.

\bibitem{kruggel2008study}
H.~Kruggel-Emden, S.~Wirtz, V.~Scherer, A study on tangential force laws
  applicable to the discrete element method (dem) for materials with
  viscoelastic or plastic behavior, Chemical Engineering Science 63~(6) (2008)
  1523--1541.

\bibitem{balevivcius2021experimental}
R.~Balevi{\v{c}}ius, A.~Maknickas, I.~Sielamowicz, Experimental, continuum-and
  dem-based velocities in a flat-bottomed bin, Powder Technology 377 (2021)
  297--307.

\bibitem{golshan2020experimental}
S.~Golshan, B.~Esgandari, R.~Zarghami, B.~Blais, K.~Saleh, Experimental and dem
  studies of velocity profiles and residence time distribution of non-spherical
  particles in silos, Powder Technology 373 (2020) 510--521.

\bibitem{alizadeh2013characterization}
E.~Alizadeh, O.~Dub{\'e}, F.~Bertrand, J.~Chaouki, Characterization of mixing
  and size segregation in a rotating drum by a particle tracking method, AIChE
  Journal 59~(6) (2013) 1894--1905.

\end{thebibliography}

\end{document}